\documentclass[aps, prd, showpacs, superscriptaddress, groupedaddress, floatfix]{revtex4-2}
\usepackage{caption}
\usepackage{graphicx}
\usepackage{subfig}
\usepackage[dvipsnames]{xcolor}
\usepackage{amsmath}
\usepackage{amssymb}
\usepackage{amsfonts}
\usepackage{amsthm}
\usepackage{mathtools}
\usepackage{mathrsfs}
\usepackage{dcolumn}
\usepackage{rotating, bm, array}
\usepackage{booktabs}
\usepackage{comment}
\newcommand{\ignore}[1]{}

\usepackage[utf8]{inputenc}
\usepackage[english]{babel}
\usepackage[normalem]{ulem}
\usepackage[pagebackref=false, colorlinks=true]{hyperref}
\hypersetup{linkcolor=blue, citecolor=blue, urlcolor=blue}

\newcommand{\dd}{\mathrm{d}}
\newcommand{\RR}{\mathcal{R}}

\begin{document}

\title{A Potential Black Hole Mimicker From Non-Minimal Coupling}

\author{Debanjan Debnath}
\email{debjanjan@gmail.com}
\affiliation{Department of Physics, Indian Institute of Technology Kanpur, Uttar Pradesh - 208016, India}

\author{Rikpratik Sengupta}
\email{rikpratiks@iitk.ac.in}
\affiliation{Department of Physics, Indian Institute of Technology Kanpur, Uttar Pradesh - 208016, India}

\author{Kaushik Bhattacharya}
\email{kaushikb@iitk.ac.in}
\affiliation{Department of Physics, Indian Institute of Technology Kanpur, Uttar Pradesh - 208016, India}

\begin{abstract}
We present a class of horizonless, regular ultra-compact objects arising in a theory of gravity that allows curvature-fluid coupling. The non-minimal interaction between fluid variables and the Ricci scalar generates a vacuum-like equation of state in the interior, while the exterior remains exactly Schwarzschild. The two spacetimes are glued through a thin shell at the junction. The interior metric is non-singular, the shell acquires a stiff-matter equation of state, and near-horizon compactness allows the configuration to mimic black-hole phenomenology without an event horizon. For radii in the astrophysically motivated range $5$--$7$ km, imposed as an external input rather than predicted by the theory, the model's mass-compactness relation maps this radius window onto a mass window of $1.4$--$2.1\,M_\odot$ for a suitable choice of the curvature-fluid coupling coefficient. We carry out a detailed stability analysis of the shell in the frozen-bulk approximation, identify the compactness window in which the interior is everywhere regular and the shell is stable, and quantify the accuracy of the frozen-bulk approximation through an explicit adiabaticity parameter. We then develop the surface thermodynamics of the shell assuming a first order phase transition-like origin of the ultra-compact object.
%and show that its stiff equation of state is the generic signature of a two-dimensional conformal sector living on the phase %boundary between the non-minimally coupled interior and the general relativistic exterior. 
This idea then gives rise to shell temperature and radiation, the shell temperature scaling $T_\Sigma \propto M^{-1/2}$ 
%\uline{and the mass independence of the shell luminosity} 
turns out to be a falsifiable predictions of the model.
\end{abstract}

\maketitle

\section{Introduction}
\label{intro}

The classical description of gravitational collapse in general relativity (GR) generically culminates in the formation of spacetime singularities, where curvature invariants, energy densities, and tidal forces diverge without bound. This is one of the oldest and least comfortable predictions of the theory: a point at which the equations that describe gravity simply stop making sense. Since Sakharov's seminal idea of induced gravity~\cite{Sakharov1967}, it has long been expected that vacuum polarization and quantum matter effects in curved spacetime can generate effective non-minimal interactions between geometry and matter fields, and that these interactions become important precisely in the strong-field regime where classical GR predicts a singularity. In strong gravity regimes, such couplings naturally modify the effective stress-energy sector and may dynamically generate repulsive gravitational effects capable of halting collapse before singularity formation~\cite{GarciaLobo2010}. From this perspective, curvature-coupled matter theories provide a physically motivated semiclassical framework for exploring regular ultra-compact configurations without abandoning the geometric foundations of GR, in contrast to approaches that modify the gravitational action itself at a fundamental level.

A concrete route to this kind of behavior can be traced to the physics of dark energy. An overdensity of dynamical, clustered dark energy may undergo gravitational collapse \cite{Basse:2010qp}, and if the fluid supports a non-minimal coupling to curvature, this collapse can trigger the coupling once the local fluid energy density crosses a threshold value. In the final stages of such a collapse the non-minimally coupled state may cease to be energetically favored relative to ordinary general relativity, giving rise to a phase-separated configuration in which the old GR phase reappears and expands outward. The non-minimally coupled phase then survives only inside a bubble embedded in the GR phase, and the bubble wall behaves precisely like a domain wall separating two regions governed by different effective gravitational dynamics. This is the physical picture underlying the construction developed in this paper.

In this paper, we show explicitly that the interior spacetime, containing a curvature-coupled dark energy-like fluid, can be glued to an external Schwarzschild solution through a thin shell that acts as this domain wall. The shell separates two spacetime regions in which the effective gravitational field equations differ, and its mechanical equilibrium is fixed self-consistently by the gravitational junction conditions alone, within the frozen-bulk stability analysis of Sec.~\ref{sec:stability}. Once this mechanical equilibrium is combined with the surface thermodynamics of Sec.~\ref{sec:thermo}, it fixes a specific domain wall temperature, $T_\Sigma$, and in the near-horizon, ultra-compact limit this temperature is directly related to the mass $M$ of the exterior Schwarzschild spacetime. Mass-temperature relations of one kind or another are common throughout astrophysics, but the shell temperature obtained here follows a distinctive high-compactness scaling, $T_\Sigma \propto M^{-1/2}$ at fixed compactness. This is unlike the Hawking temperature both in its numerical dependence on mass and, more importantly, in its physical origin: it arises from the geometric and thermodynamic structure of the shell-material rather than from quantum field theory in curved spacetime evaluated on a horizon, and there is in fact no horizon anywhere in the spacetime we construct. For a particular choice of the non-minimal coupling constant and for radii externally restricted to the range 5--7 km, the resulting ultra-compact object is compatible with a mass in the range $1.4$--$2.1\,M_\odot$, with a luminosity that, at fixed compactness, is independent of the mass. Because the shell radius satisfies $2M < R_\Sigma < 3M$, the configuration is a natural black hole mimicker: compact enough to reproduce essentially all of the phenomenology associated with a horizon, while possessing none.

The possibility that strong-gravity systems may avoid the formation of spacetime singularities has motivated a wide range of investigations extending beyond the classical framework of general relativity. These issues have stimulated considerable interest in regular compact objects, in quantum modifications of gravitational collapse, and in observational probes capable of distinguishing black holes from horizonless alternatives \cite{Bardeen1968,Abedi2017,CarballoRubio2018}. A major development along a conceptually different but physically similar line was the proposal of gravitational vacuum condensate stars, or gravastars, by Mazur and Mottola~\cite{Mazur2001}. Although interesting and attracting contemporary theoretical and observational attention \cite{Nakao2023, Adler2024, Rosa2024, Cardoso2016, ChirentiRezzolla2007, Sengupta2020}, the gravastar paradigm relies on several exotic ingredients whose microscopic origin remains uncertain, including the existence of a gravitational condensate phase, sharply localized phase transitions, and an effectively imposed de Sitter interior equation of state. Gravastar models are typically constructed by hand, patching together a de Sitter interior, a thin shell of stiff matter, and a Schwarzschild exterior, without a dynamical explanation of why the interior should take a de Sitter form or why the transition should occur where it does. In contrast, the framework considered here achieves qualitatively similar ultra-compact horizonless configurations through the non-minimal coupling of curvature and dark energy, so that the vacuum-like interior equation of state we find is a genuine consequence of the field equations of the theory rather than an input assumption. The present framework further connects collapse dynamics with equilibrium thermodynamics by interpreting the shell as a phase boundary whose mechanical equilibrium follows from the gravitational junction conditions alone, without any additional phenomenological restoring force, and we show that this equilibrium persists over a compactness window extending just below the Buchdahl limit \cite{Buchdahl:1959zz}, offering a coherent mechanism for black-hole mimicking compact objects that remain horizonless yet observationally testable through gravitational-wave echoes, electromagnetic signatures, and deviations in ringdown spectra.

It is worth setting out in some detail exactly how the present construction relates to, and differs from the gravastar proposal, since on the surface the two pictures look similar: both replace the interior of a would-be black hole with a de Sitter-like core and both use a thin shell to join this core to a Schwarzschild exterior. In the original Mazur-Mottola construction \cite{MazurMottola2004}, and in the closely related dark-energy-star picture of Chapline, Hohlfeld, Laughlin and Santiago \cite{Chapline2003}, the interior equation of state $p=-\rho$ is not derived from any field equation; it is imposed as a hypothesis about the end state of gravitational collapse, motivated by the analogy with a quantum phase transition or a gravitational Bose-Einstein condensate forming as the local temperature associated with the collapsing matter approaches the Unruh temperature seen by a freely falling observer near the would-be horizon. The intervening shell in this picture is likewise assigned an equation of state, usually taken to be as stiff as causality allows, essentially by hand, in order to satisfy the Israel junction conditions once the interior and exterior equations of state have already been fixed. Because the vacuum-like phase and the stiff shell are both postulated rather than derived, a great deal of subsequent work has focused on testing whether such a five-layer, or in simplified treatments three-layer, configuration is even mathematically and physically consistent: Visser and Wiltshire's original linearized stability analysis \cite{Visser2004} showed that thin-shell gravastars can be stable for a suitable choice of shell equation of state, Cattoen, Faber and Visser \cite{CattoenFaberVisser2005} argued that a strictly isotropic gravastar of this kind cannot avoid a surface layer with a singular energy density and proposed continuous-pressure generalizations to smooth this out, Horvat, Ili\'{c} and Marunovi\'{c} \cite{HorvatIlijicMarunovic2011} examined the radial stability of such anisotropic configurations in more detail, and Chirenti and Rezzolla \cite{ChirentiRezzolla2007} identified characteristic axial and polar $w$-mode gravitational-wave signatures by which an oscillating gravastar could in principle be distinguished from a black hole of the same mass, a program continued by Pani and collaborators \cite{PaniGravastarGW2009} through a full non-radial perturbation analysis of thin-shell gravastars and their gravitational-wave scattering properties.

The framework developed in this paper differs from this picture in a way that is more than cosmetic. Here the vacuum-like interior equation of state is not an input; it is the unique consequence of solving the non-minimally coupled field equations, Eq.~\eqref{f_e}, for the simplest possible choice of conformal function, Eq.~\eqref{Fcansatz}, as shown explicitly in Sec.~\ref{sec:eosn}. Likewise the shell is not assigned a stiff equation of state by fiat: we show in Sec.~\ref{sec:matchingSch} that the shell equation of state parameter $\omega_\Sigma$ is a calculable function of the coupling parameter $\rho_0$ and the compactness, and that it is driven toward the stiff, causally maximal value only in the near-horizon limit, becoming exactly stiff for the particular choice $\rho_0R^2=1/4$ rather than being postulated from the outset. The instability and singular-shell issues that motivate much of the gravastar stability literature above are, in the present framework, replaced by the concrete, quantitative diagnostics of Secs.~\ref{sec:stability} and \ref{sec:adiabaticity}: an explicit compactness window in which the interior particle number density is everywhere positive and the frozen-bulk equilibrium is stable, together with an honest adiabaticity parameter that tells us precisely how much to trust that equilibrium calculation. Finally, where the gravastar literature has generally treated the shell's thermodynamic behavior, when it is discussed at all, as an independent phenomenological addition to the geometric construction, often invoked mainly to argue that gravastars are cold and cannot radiate significantly, thereby sidestepping the information paradox, the present paper derives a genuine surface thermodynamics for the shell directly from its equation of state and some other simple assumptions.
%shows in Sec.~\ref{sec:cft} that this thermodynamics matches, term by term, the universal thermal behavior of a two-dimensional %conformal sector. 
This lets us assign a definite, non-Hawking temperature-mass law and, at fixed compactness, a mass-independent luminosity to the object, rather than treating thermal emission as either absent by assumption or as an unconstrained free function of the model.

The remainder of the paper is organized as follows. In Sec.~\ref{sec:jc} we set up the non-minimally coupled fluid action, derive the modified field equations, and work out the   junction conditions that must hold across a thin shell in this theory. In Sec.~\ref{sec:interior} we construct an explicit, exactly solvable, everywhere-regular interior solution with a vacuum-like equation of state. In Sec.~\ref{sec:matching} this interior is matched to an exterior Schwarzschild spacetime through the shell, and the shell stress-energy is worked out explicitly, including the near-horizon, gravastar-like limit. In Sec.~\ref{sec:stability} we carry out a detailed stability analysis of the shell within a frozen-bulk approximation, identify the compactness window in which the configuration is regular and stable, and quantify the size of the frozen-bulk approximation through an adiabaticity parameter built from the interior and shell curvature scales. In Sec.~\ref{sec:thermo} we develop the thermodynamics of the stiff shell, treating the interior-to-exterior transition as a first-order phase transition.
%, and show that the resulting free energy is precisely reproduced by a two-dimensional conformal field theory living on the %shell history, which lets us convert the postulated temperature-mass scaling and mass-independent luminosity into genuine %predictions in terms of a small effective central charge. 
Section~\ref{sec:discussion} discusses the physical implications of these results and the directions in which the present analysis could be extended.
%, and the Appendix collects the extrinsic curvature formulas used throughout.

%%%%%%%%%%%%%%%%%%%%%%%%%%%%%%%%%%%%%%%%%%%%%%%%%%%%%%%%%%%%%%%%%%%%%%%%%%%%%%%%%%%%%%%%%%%%%%%%
\section{Field Equations and Junction Conditions}
\label{sec:jc}

The action of a perfect fluid in general curved space which is non-minimally coupled with the Ricci scalar can be written as
\begin{align}
    \mathscr{A}_c = \frac{1}{2 \kappa} \int d^4x \sqrt{-g} \hspace{1mm}\left[1 + \alpha_c F_c(n, s) \right] R + \mathscr{A}_{\text{fluid}} + \mathscr{A}_\Sigma.
\label{beqnnm}
\end{align}
Here $F_c$ is a function of the particle number density $n$ and entropy per particle $s$; $\alpha_c$ is the coupling constant, $R$ is the Ricci scalar, $\mathscr{A}_\Sigma$ is the action corresponding to the possibly formed shell and $\mathscr{A}_{\text{fluid}}$ is the action functional for the fluid.

The general Bettoni-Liberati construction on which Eq.~\eqref{beqnnm} is based~\cite{bettoni2} admits, alongside the conformal coupling $F_c(n,s)R$ studied here, a disformal sector built from $F_c^{\mu\nu}(n,s)R_{\mu\nu}$, which acts on tensor rather than scalar curvature and is not simply related to the conformal term by a field redefinition. We restrict attention to the conformal sector throughout this paper because it is the coupling that preserves the perfect-fluid form of the stress tensor, Eq.~\eqref{1stpov}, exactly, so that the interior remains, in form, an ordinary general-relativistic fluid spacetime with a curvature-induced equation of state; the disformal sector generically introduces anisotropic stresses already at the level of the bulk field equations, which would qualitatively change the interior solution of Sec.~\ref{sec:interior} and the shell stress tensor of Sec.~\ref{sec:matching}. The construction presented here should accordingly be read as a controlled study of the conformal sector in isolation, motivated by its comparative analytical tractability and its preservation of fluid-like matter, rather than as a claim that the conformal interaction dominates over the disformal one on physical grounds; a treatment including the disformal term, which could modify both the compactness window of Sec.~\ref{sec:stability} and the shell thermodynamics of Sec.~\ref{sec:thermo}, is left for future work. The action for the fluid in our case is:
\begin{align}
  \mathscr{A}_{\text{fluid}} = \int d^4x \sqrt{-g} \hspace{1mm} \left[\vphantom{\alpha^{A}_{\hspace{1mm}, \hspace{0.3mm} \mu}} F(n, s) + J^{\mu} \left(\psi_{, \hspace{0.3mm} \mu} + s \theta_{, \hspace{0.3mm} \mu}  + \beta_A \alpha^{A}_{\hspace{1mm}, \hspace{0.3mm} \mu} \right) \right]. \label{Sfluid}
\end{align}
In the above, $F$ is a function of the particle number density $n$ of the constituent matter-fluid, entropy per particle $s$, and $J^{\mu}$ is the densitized particle number flux vector $\sqrt{-g}\, n U^{\mu}$, with $U^{\mu}$ the 4-velocity of a fluid particle, normalized as $U^{\mu}U_{\mu} = -1$. The action is a functional of the metric $g_{\mu\nu}$, $J^{\mu}$, the Lagrangian coordinates $\alpha^A$, and the spacetime scalars (Lagrange multipliers) $\psi$, $\theta$ and $\beta_A$. These multipliers enforce the constraint equations satisfied by the fluid, namely conservation of particle number, $J^{\mu}_{\;\, , \hspace{0.3mm} \mu} = 0$, and the entropy exchange constraint, $\left(sJ^{\mu} \right)_{, \hspace{0.3mm} \mu} = 0$.

Varying this action with respect to the metric components $g_{\mu \nu}$ gives the field equation \cite{bettoni2} as:  
\begin{align}
    \frac{1}{\kappa} (1 + \alpha_c F_c) G_{\mu \nu} = g_{\mu \nu} F - \widetilde{h}_{\mu \nu} \left(\frac{\partial F}{\partial n} + \frac{\alpha_c}{2\kappa}  \frac{\partial F_c}{\partial n} R \right) n - \frac{\alpha_c}{\kappa} \left( g_{\mu \nu} \Box F_c - \nabla_{\mu} \nabla_{\nu} F_c \right), \label{f_e}
\end{align}
where $\widetilde{h}_{\mu \nu} = g_{\mu \nu} + U_{\mu} U_{\nu}$ is the transverse metric, $G_{\mu \nu} \equiv R_{\mu \nu} - \frac{1}{2} g_{\mu \nu} R$ is the Einstein tensor, and $\Box \equiv g^{\mu \nu} \nabla_{\mu} \nabla_{\nu}$. It is useful to define an effective gravitational constant $\kappa_{\text{eff}}$ and an effective stress-energy tensor, writing Eq.~\eqref{f_e} as
\begin{align}
    \frac{1}{\kappa_{\text{eff}}}  G_{\mu \nu} =  T^{\left(\text{eff}\right)}_{\mu \nu}, \label{f_e_1}
\end{align}
with
\begin{align}
   T^{\left( \text{eff} \right)}_{\mu \nu} \equiv g_{\mu \nu} F - h_{\mu \nu} \left(\frac{\partial F}{\partial n} + \frac{\alpha_c}{2\kappa}  \frac{\partial F_c}{\partial n} R \right) n - \frac{\alpha_c}{\kappa} \left( g_{\mu \nu} \Box F_c - \nabla_{\mu} \nabla_{\nu} F_c \right),
\qquad
 \kappa_{\text{eff}} \equiv \kappa \left(1 + \alpha_c F_c \right)^{-1}.
 \label{keff}
\end{align}
For a barotropic fluid we identify the energy density and pressure in the usual way,
\begin{align}
    \rho \equiv - F, \qquad p \equiv F - n \frac{\partial F}{\partial n}. \label{1stpov}
\end{align}
In this formalism all the fluid properties are obtained from suitable variations of the fluid action.

We derive the junction conditions following the symbols and notations of E. Poisson [Sec.~3.7 of Ref.~\cite{PoissonToolkit}]. We will consider a three-dimensional hypersurface $\Sigma$ separates the whole four-dimensional spacetime into two regions, $\mathscr{V}^+$ and $\mathscr{V}^-$, with metrics $g^{+}_{\mu \nu}$ and $g_{\mu \nu}^-$ with associated coordinate-charts $x^\mu_+$ and $x^\mu_-$ respectively. Our objective is to find the constraint equations that must be imposed on the metric components and other physical quantities so that $\mathscr{V}^+$ and $\mathscr{V}^-$ are joined smoothly across $\Sigma$.

On the matching hypersurface $\Sigma$, we install coordinates $\{y^a\}$, $a = 1,2,3$, with $y^a \equiv \{\tau, \vartheta, \phi \}$, and tangent vectors $e^\mu_a := \left(\partial x^\mu/\partial y^a \right)$. The hypersurface is pierced orthogonally by a congruence of geodesics, and we let $\ell(x^\mu)$ denote proper length or proper time along these geodesics, so that the unit normal can be written $n_{\mu} = \epsilon \partial_{\mu}\ell$, with $dx^\mu = n^\mu d\ell$ and $\epsilon \equiv n^\mu n_\mu = \mp 1$ for a spacelike or timelike hypersurface. Here $\{x^\mu\}$ is a  coordinate-chart that overlaps with $\{x^\mu_-\}$ and $\{x^\mu_+\}$ within $\mathscr{V}^-$ and $\mathscr{V}^+$ respectively.  We assign $\ell < 0$ in $\mathscr{V}^-$, $\ell > 0$ in $\mathscr{V}^+$, and $\ell = 0$ on $\Sigma$; by construction $n_\mu e^\mu_a = 0$. Introducing the Heaviside function $\Theta(\ell)$, with $\Theta^2(\ell) = 1$, $\Theta(\ell)\Theta(-\ell) = 0$ and $d\Theta(\ell)/d\ell = \delta(\ell)$, we write the metric as $g_{\mu \nu} = \Theta(\ell) g_{\mu \nu}^+ + \Theta(-\ell) g_{\mu \nu}^-$, so that
\begin{align}
    \partial_\lambda g_{\mu \nu} = \Theta (\ell) \, \partial_\lambda g_{\mu \nu}^+ + \Theta (-\ell) \, \partial_\lambda g_{\mu \nu}^- + \epsilon \delta (\ell) \left[ g_{\mu \nu} \right] n_\lambda,
\end{align}
where $[X] \equiv X^+|_\Sigma - X^-|_\Sigma$ for any tensorial quantity $X$. To avoid the ill-defined product $\Theta(\ell)\delta(\ell)$ in the Christoffel symbols, we must impose continuity of the induced metric, $[g_{\mu\nu}] = 0$, i.e., $g^+_{\mu \nu} = g^-_{\mu \nu}$ on $\Sigma$. With this condition the Christoffel symbols and their derivatives take the distributional form
\begin{align}
     \Gamma^\lambda_{ \mu \nu} = \Theta (\ell) \, \Gamma^{+\lambda}_{\mu \nu} + \Theta (-\ell) \, \Gamma^{-\lambda}_{\mu \nu}, \qquad
     \partial_\lambda\Gamma^\rho_{\mu \nu} = \Theta (\ell) \, \partial_\lambda\Gamma^{+\rho}_{\mu \nu} + \Theta (-\ell) \, \partial_\lambda\Gamma^{-\rho}_{\mu \nu} + \epsilon \, \delta (\ell) \,\left[\Gamma^\rho_{ \mu\nu} \right] \,n_\lambda,
\end{align}
which in turn give the Riemann tensor and its own $\delta$-function contribution,
\begin{align}
    R^{\rho}_{\,\, \sigma \mu \nu} = \Theta (\ell) R^{+\rho}_{\,\, \sigma \mu \nu} + \Theta (-\ell) R^{-\rho}_{\,\, \sigma \mu \nu} + \delta (\ell) \, A^{\rho}_{\,\, \sigma \mu \nu},
    \qquad
    A^{\rho}_{\,\, \sigma \mu \nu} \equiv \epsilon \left(n_\mu \left[\Gamma^{\rho}_{\,\, \sigma \nu} \right] - n_\nu \left[\Gamma^{\rho}_{\,\, \sigma \mu} \right] \right). \label{Riemann_tensor_full_dist_form}
\end{align}
Tracing and contracting with $g^{\mu\nu}$ gives the corresponding singular parts of the Ricci tensor and the Ricci scalar,
\begin{align}
    R_{\mu \nu} = \Theta (\ell) R^+_{ \mu \nu} + \Theta (-\ell) R^{-}_{ \mu \nu} +  \, \delta (\ell) \, A_{\mu \nu},
\qquad
    R = \Theta (\ell) \, R^+ + \Theta (-\ell) \, R^{-} +  \delta (\ell) \, A, \label{Riemann_Dist_form}
\end{align}
with $A_{\mu \nu} \equiv A^{\rho}_{\,\, \mu \rho \nu}$ and $A \equiv g^{\mu\nu}A_{\mu\nu}$, and the singular part of the Einstein tensor follows as
\begin{align}
    \mathscr{G}_{\mu \nu} \equiv A_{\mu \nu} - \frac{1}{2} A g_{\mu \nu}. \label{ETdelta}
\end{align}
The conformal function $F_c$ must also be handled distributionally, $F_c(n) = F_c(n_+)\Theta(\ell) + F_c(n_-)\Theta(-\ell)$. Because Eq.~\eqref{f_e} contains second derivative of $F_c$, we must impose continuity of $F_c$ across $\Sigma$,
\begin{align}
     F^+_c (n_+) = F^-_c (n_-), \quad \text{on} \,\, \Sigma,
\end{align}
in order to avoid terms involving the derivative of the Dirac distribution. Taking the covariant derivative twice and tracing gives $\Box F_c(n) = \Box F_c(n_+)\Theta(\ell) + \Box F_c(n_-)\Theta(-\ell) + \epsilon[\partial_\ell F_c]\delta(\ell)$, where $n^\mu \partial_\mu F_c \equiv \partial_\ell F_c$. The last term in parentheses of Eq.~\eqref{f_e} then contributes a piece, $-(\epsilon\alpha_c/\kappa) \left(g_{\mu\nu}[\partial_\ell F_c] - n_\mu[\nabla_\nu F_c]\right)$, which is not manifestly symmetric in its indices. Symmetry is restored if we assume that the discontinuity of the first derivative of $F_c$ lies purely along the normal direction, i.e., imposing
\begin{align}
     \left[\partial_\mu F_c \right] = \epsilon n_\mu \left[\partial_\ell F_c \right],
 \end{align}
in which case the offending term becomes $-(\epsilon\alpha_c/\kappa)(g_{\mu\nu} - \epsilon n_\mu n_\nu)[\partial_\ell F_c]$, which is symmetric. A further subtlety concerns the second term on the right-hand side of Eq.~\eqref{f_e}, which involves $\partial F_c/\partial n$, the Ricci scalar $R$, and $n$ itself. Since $n = n_+\Theta(\ell) + n_-\Theta(-\ell)$ carries no $\delta$-function piece, this term contributes no singular part provided the Ricci scalar's own singular piece, which from Eqs.~\eqref{Riemann_Dist_form} and \eqref{ETdelta} can be written as (Appendix \ref{Singular_part_Ricci_scalar}): 
\begin{align}
    R = R_+ \Theta (\ell) + R_- \Theta (-\ell) - 2 \epsilon \left[K \right] \delta (\ell), \label{RicciScalar_dist}
\end{align}
where $K_{ab} \equiv n_{\alpha ; \beta} \, e^\alpha_a e^\beta_b $ is the \textit{extrinsic curvature} tensor or the \textit{second fundamental form} of the hypersurface $\Sigma$, and $\left[K\right] = h^{ab} \left[K_{ab}\right]$, vanishes identically, i.e.,
\begin{align}
    \left[K \right] = 0. \label{JC3}
\end{align}
This condition, absent in ordinary GR, arises naturally in $f(R)$-type theories whenever the field equations involve the Ricci scalar explicitly. Equating the remaining singular terms on both sides of Eq.~\eqref{f_e} then gives
\begin{align}
    \frac{1}{\kappa} (1 + \alpha_c F_c) \mathscr{G}_{\mu \nu} = S_{\mu \nu}   - \frac{\epsilon \alpha_c}{\kappa} \left(g_{\mu \nu} - \epsilon n_\mu n_\nu \right) \left[\partial_\ell F_c \right], \label{ShellSET}
\end{align}
where $S_{\mu \nu}$ is the stress-energy tensor of the shell, and we have implicitly assumed that the transverse metric $h_{\mu \nu}$ is continuous across the junction. It can be checked by direct calculation that $n^\mu \mathscr{G}_{\mu \nu} = 0 = \mathscr{G}_{\mu \nu} n^\nu$, so the shell SET admits the decomposition $S^{\mu \nu} = S^{ab} e^\mu_a e^\nu_b$, and Eq.~\eqref{ShellSET} becomes
\begin{align}
    S_{ab} = -  \frac{\epsilon}{\kappa} (1 + \alpha_c F_c) \left[K_{ab} \right] + \frac{\epsilon \alpha_c}{\kappa}  \left[\partial_\ell F_c \right] h_{ab}, \label{ShellSET2}
\end{align}
with $h_{ab} = g_{\mu \nu} e^\mu_a e^\nu_b$, and $\mathscr{G}_{\mu \nu} e^\mu_a e^\nu_b = -\epsilon\left(\left[K_{ab} \right] - \left[K \right] h_{ab} \right) = -\epsilon \left[K_{ab} \right]$ having been used. Equation~\eqref{ShellSET2} is the central result of this section: it relates the surface SET of the thin shell to the jump in extrinsic curvature across the hypersurface and to the discontinuity of the normal gradient of the conformal function. The relevant extrinsic curvature formulas are collected in the Appendix.

It is worth pausing to note what kind of gravitational theory Eq.~\eqref{f_e} actually describes, since this shapes every subsequent result. Writing the field equation in the effective form of Eq.~\eqref{f_e_1}--\eqref{keff}, the coupling factor $(1+\alpha_cF_c)^{-1}$ plays the role of a matter-dependent, spatially varying effective Newton constant, in much the same spirit as the scalar-tensor theories descended from Brans-Dicke gravity, except that here the role usually played by an independent scalar field is instead played directly by a functional of the fluid variables themselves. This is a genuinely different route to modifying the strength of gravity from the one taken in the Bertolami-B\"{o}hmer-Harko-Lobo family of non-minimal curvature-matter coupling theories \cite{Bertolami2007}, where the coupling appears asymmetrically between the matter Lagrangian and the curvature and generically produces an extra, non-geodesic force on freely falling test particles, a feature that has made such theories difficult to reconcile with the equivalence principle. In the Bettoni-Liberati formulation adopted here \cite{bettoni2}, by contrast, the coupling is built directly into the variational structure of the perfect-fluid action, so that the resulting stress-energy retains the perfect-fluid form of Eq.~\eqref{1stpov} at the price of an equation of state that depends on the local curvature rather than being fixed once and for all; there is consequently no extra force of this kind, and the entire effect of the non-minimal coupling is channeled into renormalizing the effective energy density and pressure of the fluid itself. It is precisely this channeling that allows the interior solution of Sec.~\ref{sec:interior} to remain, in form, an ordinary general-relativistic perfect-fluid spacetime, even though its equation of state has a purely curvature-induced origin.

The appearance of the extra matching condition $\left[K\right]=0$ in Eq.~\eqref{JC3} deserves a similar comment, since nothing analogous to it appears in the Israel formalism for ordinary general relativity. In GR the trace of the field equations is an algebraic relation between the Ricci scalar and the trace of the stress-energy tensor, so no independent matching condition on $R$ is required beyond the usual continuity of the induced metric and the specification of the surface stress-energy through the jump in extrinsic curvature. The moment the field equations involve the Ricci scalar non-linearly, or through a coupling to an independent functional of the matter fields as in Eq.~\eqref{f_e}, the trace equation yields a new condition,
%becomes a genuine second-order equation for a new propagating degree of freedom, 
physically analogous to the scalar field of Brans-Dicke theory or to the extra scalaron of $f(R)$ gravity, although in our case the fluid coupling function is not an independent scalar field.
%Matching the scalar degree of freedom across an interface generically requires continuity of both the field and its normal %derivative if the interface is not itself to source a delta-function source for that field, and Eq.~\eqref{JC3} is exactly the %geometric translation of this statement into the language of extrinsic curvature. 
In such theories additional continuity or regularity conditions are often required at interfaces in order to control singular contributions to the field equations. From this broader perspective, Eq.~\eqref{JC3} may be viewed as the geometric manifestation of the additional matching structure induced by the curvature-fluid coupling. The same structural feature is well known to arise in the thin-shell formalism for Gauss-Bonnet and more general Lovelock theories of gravity, where additional matching conditions beyond the Israel ones are similarly required whenever the underlying action departs from the strictly linear-in-curvature Einstein-Hilbert form.

%%%%%%%%%%%%%%%%%%%%%%%%%%%%%%%%%%%%%%%%%%%%%%%%%%%%%%%%%%%%%%%%%%%%%%%%%%%%%%%%%%%%%%%%%%%%%%%%%%%%%%%%%%%%%%%%%
\section{Regular Solution With Vanishing Conformal Function at the Boundary}
\label{sec:interior}

Throughout this and the following two sections we set $\kappa=1$ for algebraic simplicity, so that the field equation used below is Eq.~\eqref{f_e} with the overall factor $1/\kappa$ dropped; $\kappa$ is restored explicitly wherever a numerical estimate in physical units is required, namely in Fig.~\ref{fig:main} and in Eq.~\eqref{TsigmaM} and Eq.~\eqref{Lrad}, where the standard value $\kappa=8\pi$ is used.

In this section we solve the modified field equations, Eq.~\eqref{f_e}, for simple forms of the conformal coupling function $F_c(n,s)$. In the static case we take $F_c \equiv F_c(n)$, so that entropy plays no active role, since we do not assume heat transfer between different parts of the fluid; all fluid variables are then purely functions of $n$. Here $n=n(r)$ is a function of $r$, in a spherically symmetric spacetime, and consequently $F_c(n)$ can also be expressed as a function of $r$.
To solve the field equations we first specify $F_c$ as a function of $r$, and only afterward determine its dependence on $n$, once the radial profile of $n$ is known. With this ansatz the field equations reduce to
    \begin{align}
    \left(1 + \alpha_c F_c \right) G^0_0 &= F  - \alpha_c e^{-2\beta} \left\{ {F}_c^{\prime \prime} - \left(\beta^{\prime} - \frac{2}{r} \right) {F}_c^{ \prime} \right\}, \label{FE00_C} \\
   \left(1 + \alpha_c F_c \right) G^1_1 &= F  - n \frac{\partial F}{\partial n} - \frac{\alpha_c }{2} n \frac{\partial F_c}{\partial n} R - \alpha_c  e^{-2\beta} \left(\alpha^{\prime} + \frac{2}{r} \right) {F}_c^{ \prime},  \label{FE11_C} \\
   \left(1 + \alpha_c F_c \right) G^2_2 &= F  - n \frac{\partial F}{\partial n} -  \frac{\alpha_c }{2} n \frac{\partial F_c}{\partial n} R  -  \alpha_c e^{-2\beta} \left\{ {F}_c^{\prime \prime} + \left(\alpha^{\prime} - \beta^{\prime} + \frac{1}{r} \right) {F}_c^{ \prime} \right\}. \label{FE22_C}
\end{align}
Subtracting Eq.~\eqref{FE22_C} from Eq.~\eqref{FE11_C} and using the definitions of the Einstein tensor components gives
\begin{align}
    \left(1 + \alpha_c F_c \right)\left( \alpha^{\prime \prime} + {\alpha^{\prime}}^2 - \alpha^{\prime} \beta^{\prime} - \frac{\alpha^{\prime} + \beta^{\prime}}{r} - \frac{1 - e^{2 \beta} }{r^2} \right)  + \alpha_c  \left\{ {F}_c^{\prime \prime} - \left(\beta^{\prime} + \frac{1}{r} \right) {F}_c^{\prime} \right\} =  0, \label{diff_metcomp}
\end{align}
a coupled differential equation that cannot be solved without specifying $F_c$ together with $\alpha$ or $\beta$. Once a functional form for $\beta$ is chosen, Eq.~\eqref{diff_metcomp} becomes an ordinary differential equation for $\alpha$ for a given $F_c(r)$.

%%%%%%%%%%%%%%%%%%%%%%%%%%%%%%%%%%%%%%%%%%%%%%%%%%%%%%%%%%%%%%%%%%%%%%%%%%%%%%%%%%%%%%%%%%%%%
\subsection{Solving the Field Equations}
\label{sec:solveFE}

We solve Eq.~\eqref{diff_metcomp} for the simplest scenario, $\beta = 0$, and show that a regular interior solution on $0\le r\le R$ follows from the choice
\begin{equation}
F_c(r)=k \left(r^2-R^2 \right),
\qquad
F_c(R)=0,
\qquad
F_c'(r)=2kr,
\qquad
F_c''(r)=2k,
\label{Fcansatz}
\end{equation}
which satisfies $F_c'' - F_c^\prime/r =0$ identically. This radial profile is an ansatz, chosen as the simplest polynomial satisfying $F_c(R)=0$ and which is regular at $r=0$; it is not derived independently from a prior choice of $F_c(n)$, but, as shown in Sec.~\ref{sec:completetheory} below, it is consistent: the resulting particle-number profile $n(r)$ can be inverted to yield an explicit, single-valued function $F_c(n)$, Eq.~\eqref{Fcofn}, of which Eq.~\eqref{Fcansatz} is the on-shell restriction to the static background. With this choice Eq.~\eqref{diff_metcomp} reduces to
\begin{align}
    \alpha^{\prime \prime} + {\alpha^\prime}^2 - \frac{\alpha^\prime}{r} = 0
    \quad\Longleftrightarrow\quad
    \left(e^\alpha \right)^{\prime \prime} - \frac{\left(e^\alpha \right)^\prime}{r} = 0,
\end{align}
which integrates to $e^{\alpha (r)} = A + B r^2$, with $A, B$ as integration constants. The interior spacetime metric is therefore
\begin{equation}
\dd s^2 = - \left(A + Br^2 \right)^2 \dd t^2 + \dd r^2 + r^2 \dd\Omega^2, \quad \text{with} \quad \dd \Omega^2 \equiv \dd \vartheta^2 + \sin^2 \vartheta \dd \phi^2,
\label{interiormetric}
\end{equation}
where $\vartheta$ and $\phi$ are the polar and azimuth angle respectively. At this point, we mention that we do not assign any `$-$' sign with the metric components and other physical  quantities in this subsection. In the subsequent discussions, when we will work out the matching conditions, the `$\pm$' signs will be spelled out explicitly in the appropriate places whenever a confusion is likely to appear. We hope that no confusion will arise due to this brevity. 

The regular branch requires $A>0$ and $A + Br^2>0$ for $0\le r\le R$. The nonzero components of the Einstein tensor are $G^0_0 =0$ and $G^1_1 = G^2_2 = G^3_3 = 4B/(A + B r^2)$, and the Ricci scalar is
\begin{equation}
\RR = -\frac{12 B}{A + B r^2}. \label{RicciSol}
\end{equation}
At the center this becomes $-12B/A$, which is finite whenever $A\neq0$; hence the interior is non-singular at $r=0$.

%%%%%%%%%%%%%%%%%%%%%%%%%%%%%%%%%%%%%%%%%%%%%%%%%%%%%%%%%%%%%%%%%%%%%%%%%%%%%%%%%%%%%%%%%%%%%%%%%%%%%%%%%%%%%%%%%%%%%%
\subsection{Energy Density, Equation of State and Particle Number Density}
\label{sec:eosn}

For $\beta=0$, Eq.~\eqref{FE00_C} gives $(1+\alpha_cF_c)G^0_0 = F - \alpha_c\left(F_c'' + 2F_c'/r\right)$, and since $G^0_0=0$, substituting the values of $F_c^{\prime\prime}$ and $F_c^\prime$ yields $F=6\alpha_ck$. Therefore
\begin{equation}
\rho=-6\alpha_ck,
\qquad
p=6\alpha_ck,
\qquad\implies \qquad
p =-\rho,
\qquad
\omega=-1,
\label{fluidEoS}
\end{equation}
so that the fluid has exactly the vacuum-like equation of state expected from the dark energy component. Positivity of the energy density requires $\alpha_ck<0$; writing $\alpha_ck\equiv-\rho_0$ with $\rho_0>0$ gives $\rho =6\rho_0>0$ and $p =-6\rho_0<0$.

To find the particle number density we use the $1$-$1$ component of the field equation, Eq.~\eqref{FE11_C}, with $F'=0$, which gives
\begin{align}
    \frac{n'}{n} = \frac{6\alpha_ck\,Br}{2 B-\alpha_ck A-2\alpha_ck BR^2+3\alpha_ck Br^2}.
\end{align}
%\begin{align}
%    \frac{n'}{n} = \frac{6\alpha_ck\,br}{2b-\alpha_cka-2\alpha_ckbR^2+3\alpha_ckbr^2}.
%\end{align}
This integrates exactly to \ignore{$n(r)=n_0 \left(2b-\alpha_cka-2\alpha_ckbR^2+3\alpha_ckbr^2\right)$} $n(r)=n_0 \left(2B-\alpha_ck A-2\alpha_ck BR^2+3\alpha_ck Br^2\right)$, and on the positive-energy branch $\alpha_ck=-\rho_0$ this becomes
\begin{equation}
n(r) = n_0 \left(2B + \rho_0 A \right) + n_0\rho_0 B \left(2R^2 - 3 r^2\right) \equiv \widetilde{n}_1 - \widetilde{n}_2 r^2,  \label{nsol}
\end{equation}
%\begin{equation}
%n(r) = n_0 \left(2b + \rho_0 a \right) + n_0\rho_0 b \left(2R^2 - 3 r^2\right) \equiv \widetilde{n}_1 - \widetilde{n}_2 r^2,  \label{nsol}
%\end{equation}
with \ignore{$\widetilde{n}_1 \equiv n_0 \left(2b + \rho_0 a + 2 \rho_0 b R^2 \right)$} $\widetilde{n}_1 \equiv n_0 \left(2B + \rho_0 A + 2 \rho_0 B R^2 \right)$ and \ignore{$\widetilde{n}_2 \equiv 3 n_0 \rho_0 b$} $\widetilde{n}_2 \equiv 3 n_0 \rho_0 B$. The particle number density increases or decreases monotonically toward the boundary depending on whether the constant $B$ is negative or positive. At the boundary, it takes the value
\begin{align}
   n_b \equiv n(R) = \widetilde{n}_1 - \widetilde{n}_2 R^2 = n_0 (2 B + \rho_0 A - \rho_0 B R^2).
\end{align}
%\begin{align}
%   n_b \equiv n(R) = \widetilde{n}_1 - \widetilde{n}_2 R^2 = n_0 (2 b + \rho_0 a - \rho_0 b R^2),
%\end{align}
The positivity of $n(R)$, imposes the following lower and upper bounds:
\begin{align}
   \rho_0 > \frac{2 B}{ B R^2 - A} \quad  \text{or} \quad  \rho_0 < \frac{2 B}{ B R^2 - A}. \label{rho0Max}
\end{align}
The lower bound is when $B<0$. The upper bound is when $B>0$ and in addition $A > BR^2$. The above discussion shows that our ansatz is capable of producing the unambiguous functional form of $n(r)$.

\begin{comment}
It is also useful to record the Eulerian-type energy density, $\rho_{\text{Eul}} = -F - (\alpha_c/2)RF_c$, which on the branch $\alpha_ck=-\rho_0$ evaluates to $\rho_{\text{Eul}} = 6 \rho_0 (A + BR^2)/(A + B r^2)$, and which can be written in terms of the particle number density using Eq.~\eqref{nsol} as
\begin{align}
    F_c (n) =  \frac{3k}{ \widetilde{n}_2} \left(n_b - n \right), \qquad \rho_{\text{Eul}} = \frac{6 \rho_0 \widetilde{n}_2 \left(A + B R^2 \right)}{A \widetilde{n}_2 + B \left(n_b - n \right)}.
\end{align}
\end{comment}

It is worth situating the interior solution constructed above within the broader family of regular black hole and regular compact object proposals, since the recurrence of a vacuum-like core across very different constructions is itself physically informative. Bardeen's original regular black hole \cite{Bardeen1968}, together with the later Hayward \cite{Hayward2006} and Dymnikova \cite{Dymnikova1992} models, all resolve the classical singularity by replacing it with an effective de Sitter core whose energy density approaches a constant as $r\to0$, so that the curvature invariants remain bounded everywhere; these constructions differ  one from another mainly in the choice of interpolating stress-energy tensor used to connect this de Sitter core smoothly to the Schwarzschild exterior at large $r$, and the stress-energy required to do so is typically introduced as an effective, phenomenological source rather than derived from a specific matter action. What we find here is structurally the same qualitative resolution, an interior whose equation of state approaches $\omega =-1$ throughout the whole interior region rather than merely near $r=0$, but arrived at by a different and, in our view, more physically transparent route: the vacuum-like behavior is not assumed as a regularizing device but falls directly out of solving the field equations of Sec.~\ref{sec:jc} once a simple, everywhere-regular conformal function is chosen. In this sense the present model can be read as providing a concrete, curvature-matter-coupled field-theoretic completion of the regular black hole program, in which the otherwise ad hoc de Sitter-like core is traced back to a genuine dynamical origin, namely the collapse and subsequent curvature-triggered freezing of a clustered dark energy fluid, rather than being posited as an equation of state chosen purely for its regularizing properties.

%%%%%%%%%%%%%%%%%%%%%%%%%%%%%%%%%%%%%%%%%%%%%%%%%%%%%%%%%%%%%%%%%%%%%%%%%%%%%%%%%%%%%%%%%%%%%%%%%%%%%%%%%%%%%%%%%%
\subsection{The Complete Fluid Theory: $F_c(n)$, $F(n)$, and the Modified Euler Equation}
\label{sec:completetheory}

The construction above specifies $F_c$ as a function of $r$ along the static background; we now record the underlying functions $F_c(n)$ and $F(n)$ of which Eq.~\eqref{Fcansatz} is the on-shell image, and verify explicitly that the modified fluid Euler equation is satisfied. This is essential not only for internal consistency of the action of Eq.~\eqref{beqnnm}, but because any subsequent perturbative treatment of the fluid requires derivatives with respect to $n$ off the static background, which cannot be read off from $F_c(r)$ alone.

Inverting the particle number density profile of Eq.~\eqref{nsol}, $n(r) = \widetilde n_1 - \widetilde n_2 r^2$, for $r^2$ and substituting into $F_c(r)=k(r^2-R^2)$ gives $F_c$ as an explicit function of $n$ alone, valid off the static background wherever the same fluid element is described by a different value of $n$,
\begin{equation}
F_c(n) = \frac{k}{\widetilde n_2}\left(\widetilde n_1 - n\right) - kR^2 = \frac{k}{\widetilde n_2}\left(n_b - n\right),
\label{Fcofn}
\end{equation}
which reproduces Eq.~\eqref{Fcansatz} when $n=n(r)$ is substituted back in, and manifestly vanishes at $n=n_b\equiv n(R)$, consistent with $F_c(R)=0$. Equation~\eqref{Fcofn} is linear in $n$, so $\partial F_c/\partial n = -k/\widetilde n_2$ is a constant, and this is the derivative that must be supplied whenever Eq.~\eqref{f_e} or Eq.~\eqref{FE11_C} is evaluated off the static profile, e.g. in a perturbative treatment of the interior.

The function $F(n)$ is fixed by the field equation itself: Sec.~\ref{sec:eosn} shows that $G^0_0=0$ forces $F=6\alpha_ck=-6\rho_0$, a constant, independent of $n$. This is not an additional assumption but a  consequence of the interior solution of Eq.~\eqref{interiormetric}; consistently, $\partial F/\partial n=0$ identically, and Eq.~\eqref{1stpov} then gives $p =F=-\rho $ for every value of $n$ along the profile, not merely at a single radius. The complete fluid theory selected by this interior solution is therefore:
\begin{equation}
F(n) = -6\rho_0, \qquad F_c(n) = \frac{k}{\widetilde n_2}\left(n_b-n\right), \qquad \frac{\partial F}{\partial n}=0, \qquad \frac{\partial F_c}{\partial n} = -\frac{k}{\widetilde n_2} = \text{constant}.
\label{completetheory}
\end{equation}
We now verify the modified fluid Euler equation. For the Bettoni-Liberati fluid action of Eq.~\eqref{Sfluid}, varying with respect to the Lagrangian coordinates $\alpha^A$ and eliminating the Lagrange multipliers in favor of the fluid variables yields the standard relativistic Euler equation for the momentum $\mu \equiv \partial F/\partial n$ conjugate to $n$,
\begin{equation}
\left(\rho +p \right) a_\mu = -\widetilde h_{\mu}^{\ \nu}\,\partial_\nu p,
\label{eulereq}
\end{equation}
with $a_\mu \equiv U^\nu\nabla_\nu U_\mu$ the fluid four-acceleration; the coupling to curvature enters Eq.~\eqref{f_e} only through the metric field equation, not through this constraint, precisely because the non-minimal coupling in Eq.~\eqref{beqnnm} is built into the perfect-fluid variational structure rather than added asymmetrically to the matter Lagrangian, as already noted in Sec.~\ref{sec:jc}; there is consequently no anomalous force term beyond Eq.~\eqref{eulereq}. On the interior solution of Sec.~\ref{sec:solveFE}, Eq.~\eqref{completetheory} gives $\rho +p = -F+F = 0$ identically, and $\partial_\nu p  = \partial_\nu F = 0$ identically as well, since $F$ is a constant throughout the interior. Both sides of Eq.~\eqref{eulereq} therefore vanish separately and identically, independent of the fluid four-acceleration $a_\mu$: the modified Euler equation is satisfied trivially by the vacuum-like interior, exactly as required for consistency of the static solution constructed in Sec.~\ref{sec:solveFE}.

%%%%%%%%%%%%%%%%%%%%%%%%%%%%%%%%%%%%%%%%%%%%%%%%%%%%%%%%%%%%%%%%%%%%%%%%%%%%%%%%%%%%%%%%%%%%%%%%%%%%
\section{Matching With an Exterior Spacetime}
\label{sec:matching}
To match the spacetimes at $r=R$, where the junction hypersurface $\Sigma$ is situated, we consider a general static, spherically symmetric spacetime of the form:
\begin{equation}
\dd s_\pm^2  = - e^{2 \alpha_\pm} \dd t_\pm^2 + e^{2 \beta_\pm} \dd r_\pm^2 + r_\pm^2 \dd\Omega^2, \label{LineEl}
\end{equation}
with $\alpha_\pm, \beta_\pm$ functions of the radial coordinate only, the $+$ and $-$ signs referring to the exterior and interior spacetime respectively. The exterior is taken to be a matter-free vacuum, so $F_c^+(r_+)=0$ throughout $\mathscr{V}^+$. Matching the first fundamental form fixes $d\tau \equiv e^{\alpha_-} dt_- = e^{\alpha_+} dt_+$ and $r_+ = r_- \equiv r$ on $\Sigma$, so that a single radial coordinate can be used throughout. The normal derivative of the conformal function is then $\partial_\ell F_c^-|_R = \epsilon e^{-\beta_-}(\partial F_c/\partial r)^-|_R = 2\epsilon kR$ and $\partial_\ell F_c^+|_R=0$, giving $[\partial_\ell F_c] = -2\epsilon kR$.

The extrinsic curvature jumps entering Eq.~\eqref{JC3} follow from the general formulas of Eq.~\eqref{Kab} in the Appendix, evaluated for a static shell:
\begin{equation}
\mathscr{K}_0 \equiv \left[ K^{\tau}_{\tau} \right]
= \epsilon \left(\alpha^\prime_+ e^{- \beta_+} -  \alpha^\prime_- e^{- \beta_-} \right),
\qquad
\mathscr{K}_1 \equiv \left[K^{\vartheta}_{\vartheta} \right] = \left[K^{\phi}_{\phi} \right] = \frac{\epsilon \left(e^{- \beta_+} -   e^{- \beta_-} \right)}{R},
 \end{equation}
and the condition $[K]=0$ of Eq.~\eqref{JC3} becomes
\begin{equation}
\left[K\right]=\mathscr{K}_0 + 2 \mathscr{K}_1 = 0. \label{JC4}
\end{equation}

%%%%%%%%%%%%%%%%%%%%%%%%%%%%%%%%%%%%%%%%%%%%%%%%%%%%%%%%%%%%%%%%%%%%%%%%%%%%%%%%%%%%%%%%%%%%%%%%%%%%%%%%%%%%
\subsection{Schwarzschild Spacetime as Exterior Spacetime}
\label{sec:matchingSch}

For the Schwarzschild exterior, $e^{2 \alpha_+ (r)} = e^{- 2 \beta_+ (r)} = 1-2M/r \equiv f(r)$ and $\alpha^\prime_+ (r) = M/ \boldsymbol{(} r^2 \sqrt{f(r)} \boldsymbol{)}$, so that
\begin{equation}
\mathscr{K}_0  =
\frac{f^\prime (R)}{2\sqrt{f(R)}} - \frac{2BR}{A + B R^2}, \qquad
\mathscr{K}_1 = \frac{\sqrt{f(R)}-1}{R}.
\end{equation}
Taking the shell stress-energy tensor to be of the perfect-fluid form $S^a_b \equiv (- \sigma, \mathscr{P}, \mathscr{P})$, Eq.~\eqref{ShellSET2} gives
\begin{align}
    \sigma = \frac{2}{\kappa R} \left(1 - \sqrt{1 - \frac{2M}{R}} - \rho_0 R^2 \right), \qquad \mathscr{P} = \frac{1}{\kappa R} \left(1 - \sqrt{1 - \frac{2M}{R}}  + 2 \rho_0 R^2 \right), 
    \label{RhoPShell}
\end{align}
where the junction condition Eq.~\eqref{JC4} has additionally been used, and this condition itself implies
\begin{align}
    \frac{2 R - 3M}{2 R \sqrt{f(R)}} = \frac{A + 2 B R^2 }{A + B R^2}. \label{eq:static-Kzero-condition-gen}
\end{align}
Denoting $\eta \equiv A/B$, this relation fixes $\eta$ for a chosen $R$ and $M$ as
\begin{align}
    \eta = R^2 \left[\frac{4 R \sqrt{f(R)} - 2 R + 3 M}{2 R - 3 M - 2 R \sqrt{f(R)}} \right]. \label{eta}
\end{align}
Equation~\eqref{eta} fixes only the ratio $\eta=A / B$; the two constants are fixed individually by matching of the time coordinate normalization at the junction. Imposing continuity of the first fundamental form at $r=R$, $e^{\alpha_-}(R)=e^{\alpha_+}(R)$, i.e., $A + B R^2=\sqrt{f(R)}$, already used above to identify $d\tau$ across $\Sigma$. Together, $A=\eta B $ and $B=\sqrt{f(R)}/(R^2+\eta)$ determine $A$ and $B$ separately for any chosen $R$ and $M$.
Positivity of the shell energy density in Eq.~\eqref{RhoPShell} requires
\begin{align}
    1 - \sqrt{1 - \frac{2M}{R}} > \rho_0 R^2. \label{RhoShellPos}
\end{align}
In the near-horizon limit, where the boundary surface lies close to the Schwarzschild radius so that the exterior closely mimics a black hole, we set $R = 2M+\varepsilon$ with $\varepsilon\ll2M$, so that $f(R)\approx \varepsilon/2M\ll1$, and Eq.~\eqref{RhoShellPos} reduces to $\rho_0R^2 <1$. In this same limit Eq.~\eqref{eta} gives $\eta \approx -R^2+\mathcal{O}(\varepsilon^{1/2})$, which through Eq.~\eqref{rho0Max} yields the same bound $\rho_0R^2<1$: the positivity conditions on the particle number density and on the shell energy density are therefore simultaneously satisfiable. The shell pressure is always positive in this limit, and Eqs.~\eqref{RhoPShell} simplify to
\begin{align}
    \sigma \approx \frac{2}{\kappa R} \left(1  - \rho_0 R^2 \right), \qquad \mathscr{P} \approx \frac{1}{\kappa R} \left(1   + 2 \rho_0 R^2 \right). \label{RhoPShellGrava}
\end{align}
Defining the shell equation of state parameter $\omega_\Sigma \equiv \mathscr{P}/\sigma$, this gives $\omega_\Sigma \approx (1+2\rho_0R^2)/\{2(1-\rho_0R^2)\}$, which becomes $\omega_\Sigma \approx +1$, the gravastar-like stiff value, for the special choice $\rho_0R^2=1/4$. In this limit, in the vicinity of $\Sigma$, the metric component $g_{00}^-$ varies as $\mathcal{O}(\varepsilon)$ and, from Eq.~\eqref{RicciSol}, the Ricci scalar formally scales as $\mathcal{O}(\varepsilon^{-1/2})$; because $\varepsilon$ is never exactly zero, however, regularity of the interior solution is always maintained.

The emergence of $\omega_\Sigma\to1$ in this near-horizon limit is worth dwelling on physically, since it is not merely a mathematical curiosity of this particular model. Zel'dovich showed long ago \cite{Zeldovich1962} that demanding the speed of sound in any matter should not exceed the speed of light, a bare requirement of causality rather than of any particular microphysics, bounds the pressure of any physically sensible barotropic fluid by its energy density, $p\le\rho$, with the stiff equation of state $p=\rho$ representing the causal ceiling itself; this is the same bound that is invoked when discussing the maximally compact, causally allowed cores of neutron stars and other ultra-dense objects. It is therefore a physically satisfying, rather than merely convenient, feature of the present construction that the shell is driven toward exactly this causal ceiling as the configuration is pushed toward the compactness needed to mimic a black hole, instead of requiring an ad hoc assumption of superluminal or otherwise acausal matter to hold the shell up against collapse. The near-horizon limit considered here can thus be viewed as a concrete realization of the general expectation, familiar from Buchdahl-type compactness bounds and their generalizations to shelled and anisotropic configurations, that any static, self-gravitating body approaching the black-hole compactness limit may be supported by matter approaching the causal stiffness bound, since it is difficult for a softer equation of state to generate sufficient pressure gradient to resist collapse at such extreme compactness.

\begin{figure*}[t]
    \centering
    \subfloat[\label{CvsX}]{%
        \includegraphics[width=0.48\textwidth]{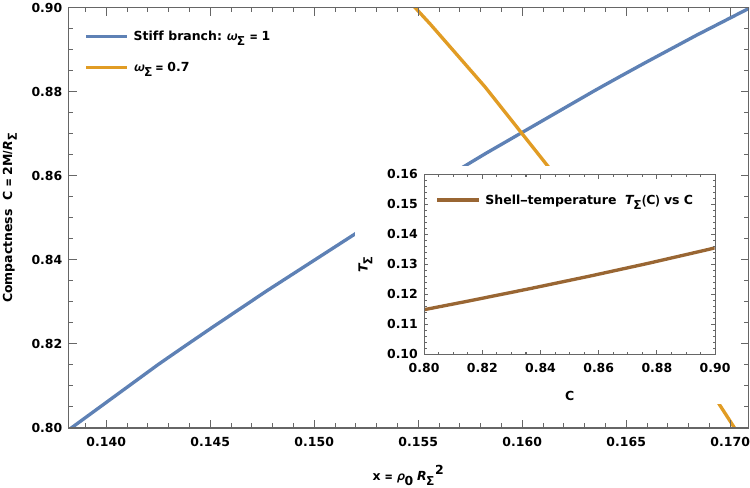}%
    }%
    \hfill
    \subfloat[\label{MvsR}]{%
        \includegraphics[width=0.48\textwidth]{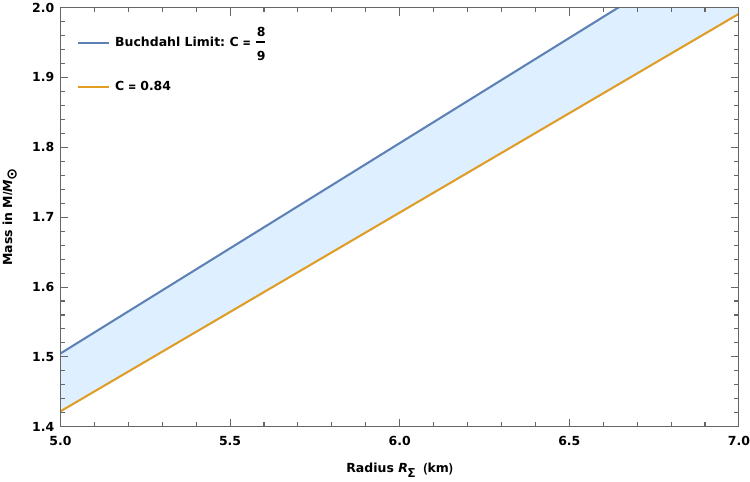}%
    }%
    \caption{The allowed parameter domain and mass-radius relations. (a) Fig.~\ref{CvsX} shows the allowed parameter domain. The orange line corresponds to an illustrative behavior for $\omega_\Sigma = 0.7$. The inset shows the variation of the shell temperature, expressed in geometrized units ($\kappa=8\pi$), with the compactness parameter. The temperature curve is plotted for the illustrative choice $M = 1,\, \gamma_\Sigma = 1$ in geometrized units, i.e. $G=c=1$ with $\kappa=8\pi$; on restoring physical units, $T_\Sigma$ has dimensions of temperature and $\gamma_\Sigma$ has dimensions of (temperature)$^{-2}\times$(energy density), consistent with $\sigma_{\mathrm{stiff}}=\gamma_\Sigma T_\Sigma^2$ in Eq.~\eqref{stiffthermo}. (b) In Fig.~\ref{MvsR} the blue strip shows the mass-radius relation restricted to the theoretically stable compactness window $0.8444\lesssim C\lesssim0.8879$ established in Sec.~\ref{sec:stability}, further restricted to the externally imposed radius range $5\,{\rm km}\le R_\Sigma\le 7\,{\rm km}$; the compactness range is a prediction of the model, while the radius range is an input rather than a prediction, so the strip should be read as the model's compactness window mapped onto an assumed radius interval, not as an independent mass-radius prediction. The strip lies just below the Buchdahl-type bound $C=8/9$ of this construction.}
    \label{fig:main}
\end{figure*}

Two clarifications on terminology are needed here. First, for the stability window established in Sec.~\ref{sec:stability},  the compactness parameter $C\equiv \frac{2M}{R_\Sigma}$ satisfies $C\lesssim0.8879$, and it corresponds to a fractional excess radius above the Schwarzschild value of
\begin{equation}
\epsilon_e \equiv \frac{R_\Sigma}{2M}-1 = \frac{1}{C}-1 \simeq 0.126\text{--}0.184\,,
\end{equation}
and a surface redshift $z_\Sigma = (1-C)^{-1/2}-1\simeq1.5$--$2$; these configurations are ultra-compact and lie well inside the photon sphere at $R_\Sigma=3M$, but their surfaces are not parametrically close to the Schwarzschild radius in the sense $\epsilon_e\ll1$, so the label ``near-horizon'' used above and elsewhere in this paper should be read as ``deep inside the photon sphere, at a compactness close to the Buchdahl bound'', not as a statement that $R_\Sigma\to2M$. Correspondingly, the exact redshift factor entering Eq.~\eqref{RhoPShell} and its neighbors, $1-2M/R_\Sigma=\epsilon_e/(1+\epsilon_e)$, may be replaced by $\epsilon_e$ only in the strict limit $\epsilon_e\ll1$, which is not the regime explored numerically in Fig.~\ref{fig:main}; where explicit numbers are quoted we use the exact expression rather than this linearized form. Second, the classic Buchdahl theorem \cite{Buchdahl:1959zz} is derived for an isotropic perfect fluid interior matched directly onto a Schwarzschild exterior with no intervening shell, under assumptions of GR that do not directly apply to the shelled, non-minimally coupled configuration considered here, which contains an additional matching condition, Eq.~\eqref{JC3}, absent from the Israel formalism.
The bound \(C<8/9\) should be understood as the endpoint of the sub-Buchdahl branch of the present construction, rather than as a direct invocation of Buchdahl's original theorem.
%The bound $C<8/9$ that recurs throughout Secs.~\ref{sec:matching}--\ref{sec:stability} should accordingly be understood as the %compactness at which the present construction's own sub-Buchdahl branch, defined by the sign of $q$ in Eq.~\eqref{qdef}, closes, %rather than as a direct invocation of Buchdahl's original theorem; the numerical coincidence with the standard GR bound is a %feature of this particular theory rather than an assumed input.

%%%%%%%%%%%%%%%%%%%%%%%%%%%%%%%%%%
%\section{Stability of the Shell}
%\label{sec:stability}

%%%%%%%%%%%%%%%%%%%%%%%%%%%%%%%%%%%%%%%%%%%%%%%%%%%%%%%%%%%%%%%%%%%%%%%%%%%%%%%%%%%%%%%%%%%%%%%%%%%%%%%%%%%%%%%%%%%%%
\section{Junction Conditions and the Frozen-Bulk Stability Window}
\label{sec:stability}

For the exterior Schwarzschild spacetime, $e^{2\alpha_+}=e^{-2\beta_+}=f(r)=1-2M/r$, and the calculation from the general formulas of the Appendix~\ref{Ext_Curv} yields:
\begin{align}
 K^{\vartheta}_{+ \vartheta} = \frac{1}{R}\sqrt{\dot R^{\,2}+f(R)},
\qquad
 K^{\tau}_{+\tau} = \frac{\ddot R + M/R^2}{\sqrt{\dot R^{\,2}+f(R)}} .
\label{Kplus}
\end{align}
Here the dot specify a derivative with the time variable $\tau$, defined on the shell. The interior regular non-minimal coupling (NMC) solution is now properly written as:
\begin{equation}
 \dd s_-^2 = - e^{2 \alpha_-}d t_-^2+ e^{-2\beta_-} d r^2+r^2 \dd\Omega^2,
\label{eq:interior-metric}
\end{equation}
where $e^{2 \alpha_-} = (A + B r^2)^2$, $e^{-2\beta_-}=1$, and $\alpha_-'(r)=2Br/(A+B r^2)$, $\beta^\prime_- = 0$, giving
\begin{align}
 K^{\vartheta}_{-\vartheta} =\frac{1}{R}\sqrt{1+\dot R^{2}},
\qquad
 K^{\tau}_{- \tau} =
 \frac{\ddot R + \dfrac{2Ba}{A+B R^2} \left(1+\dot R^{2}\right)}{\sqrt{1+\dot R^{2}}} .
\label{Kminus}
\end{align}
The trace jump is $[K]=[K^{\tau}_{\tau}]+2[K^{\vartheta}_{\vartheta}]$, and using Eqs.~\eqref{Kplus}--\eqref{Kminus} the full dynamical trace condition reads
\begin{align}
 0=[K]
 =\frac{\ddot R +M/R^2}{\sqrt{\dot R^{2}+f(R)}}
 -\frac{\ddot R +\dfrac{2B R}{A+B R^2} (1+\dot R^{2})}{\sqrt{1+\dot R^{2}}}
+\frac{2}{R}
 \left[
 \sqrt{\dot R^{2}+f(R)}-
 \sqrt{1+\dot R^{2}}
 \right] .
\label{eq:dynamic-trace-condition}
\end{align}
In the distributional NMC construction, Eq.~\eqref{eq:dynamic-trace-condition} is the exact dynamical form of the condition that removes the singular Ricci-scalar contribution. In the frozen-bulk Visser-type analysis pursued here, however, we impose this condition only on the static background, at $r = R_\Sigma$, $\dot R_\Sigma =0$, $\ddot R_\Sigma =0$, giving the algebraic background matching condition
\begin{equation}
 \frac{2R-3M}{2R\sqrt{1-2M/R}}
 =\frac{A+2BR^2}{A+BR^2} .
\label{eq:static-Kzero-condition}
\end{equation}
This fixes the ratio $A/B$ for a given compactness (see also Eqs.~\eqref{eq:static-Kzero-condition-gen} and \eqref{eta}), and the subsequent perturbative treatment should be read as a shell-probe or frozen-bulk test around a background satisfying Eq.~\eqref{eq:static-Kzero-condition}, not as a complete solution of the fully time-dependent field equations; we quantify the size of this approximation explicitly in Sec.~\ref{sec:adiabaticity}.

On the shell, $F_c(R)=0$, and for the interior profile $F_c(r)=k \left(r^2-R^2 \right)$ with $\alpha_ck=-\rho_0$, $\rho_0>0$, the dynamical normal derivative acquires a factor $\sqrt{1+\dot R^2}$, giving
\begin{equation}
 [\partial_\ell F_c] = n^\mu[\partial_\mu F_c] = -n^\mu_-\partial_\mu F_c = -2kR\sqrt{1+\dot R^2} .
\label{eq:normal-gradient-jump}
\end{equation}
The NMC junction condition, once $[K]=0$ has been imposed, is $S_{ab} = -(1/\kappa)[K_{ab}] + (\alpha_c/\kappa)[\partial_\ell F_c]h_{ab}$, and writing $S^a_b = \text{diag}(-\sigma_\Sigma, \mathscr{P}_\Sigma, \mathscr{P}_\Sigma)$, together with Eqs.~\eqref{Kplus}, \eqref{Kminus} and \eqref{eq:normal-gradient-jump}, gives the dynamical density equation
\begin{equation}
 \frac{\kappa R \sigma_\Sigma (R, \dot{R})}{2}
 =(1-\rho_0 R^2)\sqrt{1+\dot R^{2}}
 -\sqrt{\dot R^{2}+1-\frac{2M}{R}} .
\label{eq:master-density-equation}
\end{equation}
Equation~\eqref{eq:master-density-equation} is the corrected NMC-Visser master equation for the surface density. Defining, at the static radius $r=R_\Sigma$,
\begin{equation}
 %C=\frac{2M}{R_\Sigma},
 %\qquad
 y=\sqrt{1-C},
 \qquad
 x=\rho_0 R_\Sigma^2 ,
\label{eq:C-y-x-definitions}
\end{equation}
the static limit of Eq.~\eqref{eq:master-density-equation} gives $\sigma_{\Sigma_0} \equiv \sigma_\Sigma (R, \dot{R}=0)=(2/\kappa R_\Sigma)(1-y-x)$, while the static pressure, from Eq.~\eqref{RhoPShell}, is $\mathscr{P}_{\Sigma_0} = (1/\kappa R_\Sigma)(1-y+2x)$, so that the shell equation of state parameter is
\begin{equation}
 \omega_\Sigma=\frac{\mathscr{P}_{\Sigma_0}}{\sigma_{\Sigma_0}}
 =\frac{1-y+2x}{2(1-y-x)} .
\label{eq:static-omega}
\end{equation}
We invert this to write $x = (1-y)(2\omega_\Sigma -1)/\{2(1+\omega_\Sigma)\}$, and substituting back gives
\begin{align}
    \sigma_\Sigma = \frac{3(1 - y)}{\kappa R_\Sigma (1 + \omega_\Sigma)}. \label{static_sigma2}
\end{align}
For the stiff shell, $\omega_\Sigma=1$, Eq.~\eqref{eq:static-omega} implies $x=(1-y)/4$. Positivity of the shell energy density, $1-y-x>0$, written using Eq.~\eqref{static_sigma2} as $(1-y)/(1+\omega_\Sigma)>0$, holds whenever $\omega_\Sigma>-1$, since $y<1$; but positivity of $x$ itself, given the sign convention $\rho_0>0$, additionally requires
\begin{align}
    \omega_\Sigma > \frac{1}{2}.
\end{align}
Introducing
\begin{align}
    q \equiv \frac{B R^2}{A + B R^2}, \label{qdef}
\end{align}
the static trace condition, Eq.~\eqref{eq:static-Kzero-condition}, becomes $q=(3y-1)(y-1)/(4y)$. Since $A>0$ and $A+BR^2>0$, the sign of $q$ is fixed by the sign of $B$; for $y>1/3$, i.e., $C<8/9$, $q$ is negative and the branch is sub-Buchdahl, while for $y<1/3$, i.e., $C>8/9$, $q$ is positive and the branch is Buchdahl-evading. Solving for $y$ in terms of $q$,
\begin{align}
    y (q) = \frac{2 (1 + q )  - \sqrt{4 (1 + q)^2 - 3}}{3}, \label{y(q)}
\end{align}
having neglected the unphysical root, real-valuedness requires $4(1+q)^2-3\ge0$, i.e., $4q^2+8q+1\ge0$, giving $-1+\sqrt3/2 \lesssim q \le 1$, or $-0.1340\lesssim q\le1$. This translates into the overall compactness range $2/3<C<1-\left\{(4-\sqrt{13})/3\right\}^2$, i.e., $0.6667\lesssim C \lesssim 0.9827$. On the positive-$q$ branch, $0<q<1$, this becomes the Buchdahl-evading window
\begin{align}
     \frac{8}{9} < C < 1 - \left(\frac{4 - \sqrt{13}}{3} \right)^2 \;\Longrightarrow\; 0.8889 \lesssim C  \lesssim 0.9827, \label{+qBranch}
\end{align}
while the negative-$q$ range gives the sub-Buchdahl window
\begin{align}
     \frac{2}{3} < C < \frac{8}{9} \;\Longleftrightarrow\; 0.6667 \lesssim C \lesssim 0.8889. \label{-qBranch}
\end{align}
These bounds follow purely from the positivity of the shell energy density.

Assuming a constant shell equation of state $\mathscr{P}_\Sigma =\omega_\Sigma \sigma_\Sigma$, the intrinsic surface conservation equation $\dot\sigma_\Sigma + (2\dot R/R)(\sigma_\Sigma+\mathscr{P}_\Sigma)=0$ integrates to $\sigma_\Sigma(R)=\sigma_{\Sigma_0}(R_\Sigma/R)^{2(1+\omega_\Sigma)}$. Introducing $z \equiv R/R_\Sigma$, (which is distinct from $z_\Sigma$, the redshift factor) $u \equiv \dot R^2$, and substituting into Eq.~\eqref{eq:master-density-equation}, we obtain
\begin{equation}
 (1-xz^2)\sqrt{1+u}
 -\sqrt{u+1-\frac{1-y^2}{z}}
 =(1-y-x)z^{-1-2\omega_\Sigma}.
\label{eq:dimensionless-master}
\end{equation}
This can always be solved for $u$ as a function of $z$, and the equation itself takes the familiar form $\dot{R}^2 + V (R) = 0$, with $V$ acting as an effective potential; a static point is stable when $V^\prime = 0$ and $V^{\prime\prime}>0$. Instead of fixing $V$ explicitly, we rather define
\begin{align}
    S (z, u) = (1-xz^2)\sqrt{1+u}
 -\sqrt{u+1-\frac{1-y^2}{z}}
 - (1-y-x)z^{-1-2\omega_\Sigma}. \label{Sdef}
\end{align}
Since $S_{,z}+S_{,u}u^\prime(z)=0$, we have $u^\prime(z) = -S_{,z}/S_{,u}$, and similarly, using $u^\prime(z)=0$ at equilibrium, $u^{\prime\prime}(z)=-S_{,zz}/S_{,u}$. A static shell at $z=1$ satisfies $u(1)=0$, and a frozen-bulk Visser equilibrium requires $u'(1)=0$, $u^{\prime\prime}(1)<0$. Evaluating $S_{,z}$ and $S_{,u}$ at $z=1$ and substituting $x$ from Eq.~\eqref{eq:static-omega} gives
\begin{align}
    u'(1)= \frac{2 \left(\omega_\Sigma y-\omega_\Sigma +4y-1 \right)}{2\omega_\Sigma y+2\omega_\Sigma -y+2}, \label{dV}
\end{align}
so the equilibrium condition $u'(1)=0$ fixes
\begin{equation}
 y_{\rm EQ} =\frac{1+\omega_\Sigma}{4+\omega_\Sigma},
\qquad
 C_{\rm EQ}(\omega_\Sigma)
 =1-\left(\frac{1+\omega_\Sigma}{4+\omega_\Sigma}\right)^2 ,
\label{eq:y-visser-equilibrium}
\end{equation}
which for the stiff shell gives $y_{\rm EQ} = 2/5$ and $C_{\rm EQ}(1)=21/25=0.84$. Repeating the calculation for $S_{,zz}$ gives
\begin{align}
    u''(1) = - \frac{3 \left(8\omega_\Sigma^4+28\omega_\Sigma^3+24\omega_\Sigma^2-16\omega_\Sigma-29 \right)}{(1+\omega_\Sigma)^2(4\omega_\Sigma+7 )},
\end{align}
and the requirement $u''(1)<0$ yields the stability threshold
\begin{equation}
 \omega_\Sigma>\omega_{\Sigma \rm crit}\simeq0.8854,
\label{eq:omega-critical}
\end{equation}
so that the stiff shell, sitting at $\omega_\Sigma=1$, is stable in the minimal frozen-bulk Visser sense, with the corresponding compactness bound $0.8400 \le C_{\rm EQ} \lesssim 0.8511$.

\begin{figure*}[t]
    \centering
        \includegraphics[width=0.80\textwidth]{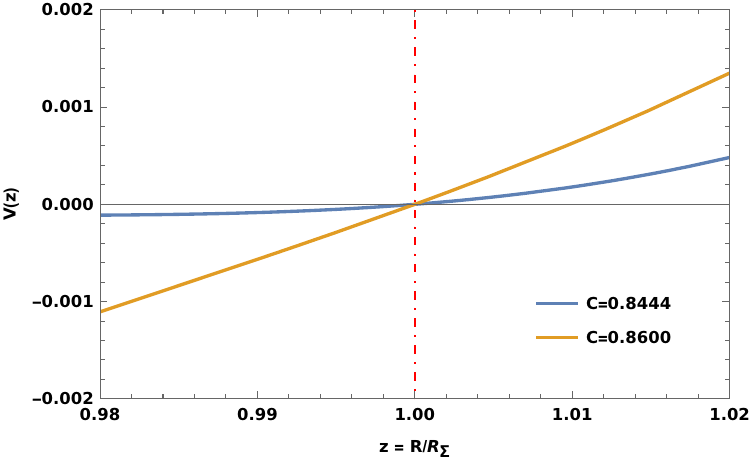}
    \caption{The dimensionless frozen-bulk Visser potential $V (z)$, computed from Eq.~\eqref{V(z)}, for $C = 0.8444$ and $C = 0.8600$. The reference shell is at $z = 1$, shown as the red dot-dashed line. The potential becomes practically flat in the lower part of the positive-density window (Eq.~\eqref{eq:C-positive-density-window}).}
     \label{fig:V(Z)vsZ}
\end{figure*}

Defining $E(z) \equiv (1-y-x)z^{-1-2\omega_\Sigma}$, $G(z) \equiv 1-(1-y^2)/z$, and $A(z) \equiv 1-xz^2$, Eq.~\eqref{eq:master-density-equation} becomes $A(z)\sqrt{1+u} - \sqrt{G(z)+u} = E(z)$, which after defining $B(z)\equiv\sqrt{1+u}$ and squaring gives the quadratic $(A^2-1)B^2 - 2AEB + E^2 - G + 1 = 0$. Solving, and choosing the root for which $B(1)=1$ (using the fact that the discriminant is a perfect square, $(1-y+xy)^2$), gives
\begin{align}
    B(z) = \frac{AE-\sqrt{E^2+(A^2-1)(G-1)}}{A^2-1}. \label{Bexpr}
\end{align}
We can define the effective potential through $\dot R^2+V(R)=0$. In terms of $z$ this equation becomes: $R_\Sigma^2 \dot{z}^2 + V(z)=0$, differentiating this equation once more with respect to $\tau$ we get:
\begin{eqnarray}
\ddot{z} + \frac{1}{2R_\Sigma^2}V^\prime(z)=0\,,
\end{eqnarray}
where the prime represents a derivative with respect to $z$. Near equilibrium $V^\prime(z_{\rm eq})=0$ and we have $V^\prime(z)\simeq V^{\prime \prime}(z_{\rm eq})\delta z$, and consequently:
\begin{eqnarray}
\ddot{\delta z} + \frac{V^{\prime \prime}(z_{\rm eq})}{2R_\Sigma^2}\delta z=0\,.
\end{eqnarray}
In the present case:
\begin{align}
    V(z) = 1 - \left[\frac{A E - \sqrt{E^2 + (A^2 -1)(G - 1)}}{A^2 - 1} \right]^2. \label{V(z)}
\end{align}
We show the variation of $V$ with $z$ in Fig.~\ref{fig:V(Z)vsZ}. Since $V(1)=0$, we can Taylor expand around $z=1$,
\begin{align}
    V(z) = V^\prime (1) (z-1) + \frac{1}{2} V^{\prime \prime} (1) (z-1)^2 + ...,
\end{align}
with coefficients
\begin{align}
    V^{\prime} (1) = \frac{2(2-5y)}{4+y}, \qquad V^{\prime \prime} (1) = \frac{2 \left(30y^3 + 269y^2 + 459y - 185 \right)}{(4+y)^3}.
    \label{dV(1)ddV(1)}
\end{align}
$V^\prime(1)$ vanishes at $y=2/5$, corresponding to $C=21/25=0.84$, slightly below the positive-density window given in Eq.~\eqref{+qBranch}. 
Within that window, $5\sqrt{13}-18<2-5y<1/3$, or $0.0277\lesssim y\lesssim0.3333$, so $V^\prime(1)$ is small and positive throughout, and an outward excursion with $z>1$ would drive $V$ positive, which is forbidden by the defining relation $\dot R^2+V(R)=0$. In the present case we see that $V^\prime(1) \ne 0$ in general, showing a light departure from the exact equilibrium analysis we have presented. We will opine on this mild departure from an exact equilibrium later.

The interior number density, $n(r)=n_0(2B+\rho_0 A)+n_0\rho_0B(2R^2-3r^2)$, can be written using $s=(r/R)^2$ and $q=BR^2/\left(A+BR^2 \right)$ as $n(s)=n_0BN(s)$, (this $s$ is not related to entropy density) with
\begin{equation}
N(s)=2+x\left(\frac{1}{q}+1-3s\right) .
\label{eq:An-definition}
\end{equation}
For the stiff branch, inserting $x=(1-y)/4$ and $q=(3y-1)(y-1)/4y$ into Eq.~\eqref{eq:An-definition} gives
\begin{align}
 N(0)
 =-\frac{3(y^2-8y+3)}{4(3y-1)},
\qquad
N(1)
 =\frac{3(y^2+2y-1)}{2(3y-1)} .
\label{eq:An-zeroone}
\end{align}
On the regular negative-$q$ branch, $B<0$, $A+BR^2>0$, and hence $q<0$, corresponding to the sub-Buchdahl condition $y>1/3$, i.e., $C<8/9$. The endpoint values in Eq.~\eqref{eq:An-zeroone} share the same negative sign in the interval
\begin{equation}
 \frac{1}{3}<y<4-\sqrt{13}
\quad\Longleftrightarrow\quad
 8\sqrt{13}-28<C<\frac{8}{9},
\label{eq:C-positive-density-window}
\end{equation}
numerically $0.8444 \lesssim C \lesssim 0.8889$. \label{eq:C-positive-density-window-numeric} In this interval $N(0)$ and $N(1)$ are both negative, and since $B<0$, choosing $n_0>0$ guarantees $n(r)>0$ for $0\le r\le R$, in addition to the regularity requirements $A>0$, $A+Br^2>0$. On this negative-$q$ branch these conditions can be satisfied, for example, by taking $B=-B_0/R^2$ with $B_0>0$ and fixing $A$ from the $q(y)$ relation. The positive-$q$, Buchdahl-evading branch of Eq.~\eqref{+qBranch} also has $N(0)$ and $N(1)$ both positive, and therefore also admits $n(r)>0$ throughout the interior; however, as shown above, this branch is not selected by the minimal constant-$\omega_\Sigma$ Visser-stable stiff point.

The exact stiff frozen-bulk Visser point, $C_{\rm EQ}(1)=0.84$, lies just below the positive-density window of Eq.~\eqref{eq:C-positive-density-window}; the mismatch,
\begin{equation}
 \Delta C
 =\left(8\sqrt{13}-28\right)-\frac{21}{25}
 \simeq 0.0044 ,
\label{eq:compactness-mismatch}
\end{equation}
is small compared with the approximations already built into the frozen-bulk, constant-EoS analysis. The curvature $V''(1)$ in Eq.~\eqref{dV(1)ddV(1)} remains positive until $30y^3+269y^2+459y-185=0$, whose relevant root is $y\simeq0.33487$, $C_{\rm curv}\simeq0.88786$, so almost the entire positive-density window carries positive Visser curvature, even though the exact equilibrium point sits slightly below it. Table~\ref{tab:mildness} gives representative values of $V'(1) = - u^\prime (1)$ and $V''(1)$ across the window.

\begin{table}[h]
\centering
\begin{tabular}{c|c|c}
$C$ & $V'(1)$ from Eq.~(\ref{dV(1)ddV(1)}) & $V''(1)$ from Eq.~\eqref{dV(1)ddV(1)} \\
\hline
$0.8444$ & $0.0126$ & $0.9367$ \\
$0.8500$ & $0.0290$ & $0.8257$ \\
$0.8600$ & $0.0591$ & $0.6207$ \\
$0.8700$ & $0.0905$ & $0.4070$ \\
$0.8800$ & $0.1233$ & $0.1834$ \\
$8/9$    & $0.1538$   & $-0.0246$
\end{tabular}
\caption{Mildness indicators for the stiff $[K]=0$ branch in the positive-density window. The drift measure $V'(1)$ is small near the lower endpoint of Eq.~\eqref{eq:C-positive-density-window}, while the curvature $V''(1)$ remains positive until very close to the Buchdahl endpoint.}
\label{tab:mildness}
\end{table}

%%%%%%%%%%%%%%%%%%%%%%%%%%%%%%%%%%%%%%%%%%%%%%%%%
The conclusion from Eqs.~\eqref{dV}--\eqref{eq:compactness-mismatch} and Table~\ref{tab:mildness} is that the lower part of the positive-density window is stable or nearly stable in the Visser sense. Although the reference configuration \(z=1\) is not an exact stationary point of the frozen-bulk potential within the positive-density window, the potential remains locally convex and its slope is small. The corresponding equilibrium is therefore only slightly displaced from \(z=1\), in the same sense that the equilibrium position of a vertically suspended massive spring is displaced from its unstretched length by the gravitational force.
These are slightly displaced frozen-equilibrium Visser points. 
%and a positive phase-boundary stiffness of the kind introduced in Sec.~\ref{sec:thermo} can naturally shift the effective %equilibrium into the physical positive-density interval. 
Combining the strict frozen-bulk equilibrium of Eq.~\eqref{eq:y-visser-equilibrium} with the positive number density window of Eq.~\eqref{eq:C-positive-density-window}, and noting that these two independent conditions nearly coincide, up to the small mismatch of Eq.~\eqref{eq:compactness-mismatch}, we identify three physically distinct regimes. The interval
\begin{equation}
 0.8444\lesssim C\lesssim0.86
\label{eq:conservative-window}
\end{equation}
is a conservative regular compactness window in which $n(r)>0$, the interior is nonsingular, and the residual Visser departure is mild, indicating a stable point slightly displaced from the potential minimum. The extended interval $0.86\lesssim C\lesssim0.8879$ remains positive-density and has positive Visser curvature, but the departure from the frozen-bulk equilibrium grows steadily larger. For $0.8879\lesssim C<8/9$ the frozen-bulk Visser curvature $V''(1)$ turns negative, so the minimal, constant-$\omega_\Sigma$ analysis pursued here no longer establishes stability; we make no stability claim in this narrow sub-window, and a fully time-dependent treatment of the coupled interior-shell system, of the kind discussed in Sec.~\ref{sec:adiabaticity}, would be needed to determine whether the configuration can be stabilized there. All stability claims made in this paper are therefore restricted to $C\lesssim0.8879$.

%%%%%%%%%%%%%%%%%%%%%%%%%%%%%%%%%%%%%%%%%%%%%%%%%%%%%%%%%%%%%%%%%%
\subsection{Validity of the frozen-bulk approximation}
\label{sec:adiabaticity}

The frozen-bulk treatment above holds the interior parameters $A,B$, and hence $q$, fixed at their static values while allowing only $R(\tau)$ to vary around the equilibrium point. This approximation is valid only if the interior geometry's own characteristic length scale is much shorter than the length scale over which the shell's radial perturbative motion is confined, so that the interior effectively relaxes instantaneously compared to the shell. We will now try to make this separation of scales explicit and quantitative.

The interior curvature scale ($l_{\rm int}$) at the shell follows directly from the Ricci scalar, Eq.~\eqref{RicciSol}, evaluated at $r=R_\Sigma$ using the definition of $q$ in Eq.~\eqref{qdef}:
\begin{equation}
\mathcal{R}^{\rm (int)}(R_\Sigma) = -\frac{12B}{A+BR_\Sigma^2} = -\frac{12q}{R_\Sigma^2},
\qquad
l_{\rm int}^2 \equiv \frac{1}{|\mathcal{R}^{\rm (int)}(R_\Sigma)|} = \frac{R_\Sigma^2}{12|q|}.
\end{equation}
The shell's own confinement scale follows from the curvature of the frozen-bulk potential, Eq.~\eqref{V(z)}, converted from the dimensionless variable $z=R/R_\Sigma$ to the physical radius via $d^2/dR^2 = R_\Sigma^{-2}\,d^2/dz^2$:
\begin{equation}
l_{\rm shell}^2 \equiv \frac{1}{\left|\dfrac{d^2 V}{dR^2}\right|_{R_\Sigma}} = \frac{R_\Sigma^2}{|V''(1)|}.
\end{equation}
The dimensionless ratio of these two length scales,
\begin{equation}
\mathcal{S} \;\equiv\; \frac{l_{\rm int}^2}{l_{\rm shell}^2} \;=\; \frac{|V''(1)|}{12|q|},
\label{adiabaticitydef}
\end{equation}
is the natural adiabaticity parameter: $\mathcal{S}\ll1$ would indicate that the interior geometry varies over scales far shorter than the region where the shell dynamics is significant, justifying the frozen-bulk treatment as a controlled leading-order approximation, while $\mathcal{S}\gtrsim1$ indicates no such separation.

Using $q(y)$ from Eq.~\eqref{qdef} and $V''(1)$ from Eq.~\eqref{dV(1)ddV(1)}, both evaluated on the stiff, negative-$q$ branch, $\mathcal{S}$ reduces to the single closed-form function of $y=\sqrt{1-C}$,
\begin{equation}
\mathcal{S}(y) \;=\; \frac{2y\left(30y^3+269y^2+459y-185\right)}{3(4+y)^3(3y-1)(1-y)}.
\label{adiabaticity}
\end{equation}
Table~\ref{tab:adiabaticity} evaluates $\mathcal{S}(y)$ at the same compactness values as Table~\ref{tab:mildness}.

\begin{table}[h]
\centering
\begin{tabular}{c c c}
$C$ & $q$ & $\mathcal{S}=|V''(1)|/12|q|$ \\
\hline
0.8444 & $-0.0704$ & $1.109$ \\
0.8500 & $-0.0640$ & $1.075$ \\
0.8600 & $-0.0512$ & $1.010$ \\
0.8700 & $-0.0362$ & $0.937$ \\
0.8800 & $-0.0185$ & $0.826$ \\
\end{tabular}
\caption{The adiabaticity parameter $\mathcal{S}$ across the positive-density, stable window. $\mathcal{S}=O(1)$ throughout.}
\label{tab:adiabaticity}
\end{table}

Across the entire positive-density, Visser-stable window, $\mathcal{S}$ remains of order unity, $0.83\lesssim\mathcal{S}\lesssim1.11$, rather than small. There is therefore no parametric hierarchy separating the interior curvature scale from the shell's confinement scale, and the frozen-bulk construction of the preceding subsections should be understood as an order-of-magnitude estimate of the equilibrium and its stability, rather than the leading term of a systematically controlled adiabatic expansion. This makes precise, rather than merely qualitative, the caveat already noted below Eq.~\eqref{eq:static-Kzero-condition}: a fully time-dependent treatment in which $A$ and $B$ are sourced self-consistently by $R(\tau)$ would be needed to promote this estimate to a controlled result, with $\mathcal{S}\to0$ the concrete condition such a treatment would need to satisfy for the present frozen-bulk analysis to be recovered as its strict limit.

The physical content of $\mathcal{S}=\mathcal{O}(1)$ is worth spelling out explicitly, since it is easy to read an order-unity adiabaticity parameter as a sign that something has gone wrong with the calculation, when in fact it is closer to a structural feature of extreme compactness itself. The logic behind treating $A$ and $B$ as frozen while $R(\tau)$ evolves is directly analogous to the Born-Oppenheimer separation used throughout physics whenever a fast, light subsystem is assumed to relax instantaneously in the potential generated by a slow, heavy one; the approximation is controlled precisely when the two associated timescales, or equivalently length scales, are widely separated, and it becomes uncontrolled precisely when they merge. A compact self-gravitating body sitting close to the Buchdahl limit is, almost by definition, a system in which every characteristic length scale of the problem, the stellar radius, the local curvature radius, and the light-crossing time of the interior, is being squeezed toward the same order of magnitude as the gravitational radius $2M$ itself; this is, after all, the entire reason such configurations are of interest as black hole mimickers in the first place. It would therefore have been surprising, rather than reassuring, to find a parametrically small adiabaticity ratio in this regime, since a genuine separation of scales is difficult to reconcile with a configuration whose defining feature is the near coincidence of all of its natural length scales. The value $\mathcal{S}=\mathcal{O}(1)$ obtained here should accordingly be read less as evidence against the frozen-bulk approach and more as a quantitative confirmation that the model has been pushed into the genuinely strong-field, near-Buchdahl regime it was designed to probe, with the associated cost that the equilibrium and stability results of this section carry the accuracy of an order-of-magnitude estimate rather than a systematically improvable expansion.

%%%%%%%%%%%%%%%%%%%%%%%%%%%%%%%%%%%%%%%%%%%%%%%%%%%%%%%%%%%%%%%%%%%%%%%%%%%%%%%%%%%%%%%%%%%%%%%%%%%%%%%%%%%%%%%%
\section{Thermodynamics of a Stiff Thin Shell at an NMC-to-GR Phase Boundary}
\label{sec:thermo}

In this section we present the thermodynamics of the shell separating the two phases. It is seen that this shell can have interesting and nontrivial thermodynamic properties which may lead to radiation from the shell.
%%%%%%%%%%%%%%%%%%%%%%%%%%%%%%%%%%%%%%%%%%%%%%%%%%%%%%%%%%%%%%%%%%%%%%%%%%%%%%
%\subsection{Surface Thermodynamics and the NMC-to-GR Phase Transition}
We consider a compact object with a regular interior described by the matter-curvature coupled theory of Sec.~\ref{sec:jc} and an exterior Schwarzschild region governed by ordinary GR. The relevant non-minimal coupling factor is $\Phi = 1 + \alpha_c F_c$, which is unity in the exterior, $\Phi_+=1$, and different from unity in the interior away from the boundary, $\Phi_-(r)\neq1$. At the phase boundary we impose the matching condition $\Phi_-(R)=1$, equivalent to $F_c(R)=0$, so that the shell is naturally interpreted as the phase boundary between an ``NMC phase,'' $\Phi\neq1$, and a ``GR phase,'' $\Phi=1$. The shell carries a surface stress tensor $S^a_b=\text{diag}(-\sigma_\Sigma,\mathscr{P}_\Sigma,\mathscr{P}_\Sigma)$, and we assume $\omega_\Sigma \equiv \mathscr{P}_\Sigma/\sigma_\Sigma \simeq 1$, the stiff-shell assumption already motivated in Sec.~\ref{sec:stability}.

Let the shell-area be $A_\Sigma = 4\pi R^2_\Sigma$. The Helmholtz free energy of a homogeneous shell is $F_\Sigma = A_\Sigma f_\Sigma(T_\Sigma)$, where $T_\Sigma$ is the shell temperature and $f_\Sigma$ the surface free-energy density per unit area. The surface entropy density is $s_\Sigma = -df_\Sigma/dT_\Sigma$, and the surface energy density is $\sigma_\Sigma = f_\Sigma + T_\Sigma s_\Sigma = f_\Sigma - T_\Sigma f'_\Sigma(T_\Sigma)$. The surface pressure is $\mathscr{P}_\Sigma = -(\partial F_\Sigma/\partial A_\Sigma)_{T_\Sigma,N_\Sigma}$, which reduces to $\mathscr{P}_\Sigma=-f_\Sigma$ when $f_\Sigma$ has no explicit area dependence, giving
\begin{equation}
\omega_\Sigma = \frac{\mathscr{P}_\Sigma}{\sigma_\Sigma} = \frac{-f_\Sigma}{f_\Sigma - T_\Sigma f'_\Sigma} .
\end{equation}

For constant $\omega_\Sigma$, the relation $-f_\Sigma=\omega_\Sigma(f_\Sigma-T_\Sigma f'_\Sigma)$ gives $f'_\Sigma/f_\Sigma = \left\{(1+\omega_\Sigma)/\omega_\Sigma\right\}(1/T_\Sigma)$, which integrates to $f_\Sigma(T_\Sigma)=f_0 T_\Sigma^{(1+\omega_\Sigma)/\omega_\Sigma}$, with $f_0$ as an integration constant. For the stiff-shell case, $\omega_\Sigma=1$, this becomes
\begin{equation}
f_{\mathrm{stiff}}(T_\Sigma) = -\gamma_\Sigma T_\Sigma^2, \qquad f_0 \equiv - \gamma_\Sigma > 0 ,
\label{fstiffdef}
\end{equation}
with the negative sign chosen so that the pressure and energy density come out positive. Indeed
\begin{equation}
s_{\mathrm{stiff}} = -\frac{df_{\mathrm{stiff}}}{dT_\Sigma} = 2\gamma_\Sigma T_\Sigma ,
\qquad
\sigma_{\mathrm{stiff}} = f_{\mathrm{stiff}} + T_\Sigma s_{\mathrm{stiff}} = \gamma_\Sigma T_\Sigma^2 ,
\qquad
\mathscr{P}_{\mathrm{stiff}} = -f_{\mathrm{stiff}} = \gamma_\Sigma T_\Sigma^2 ,
\label{stiffthermo}
\end{equation}
so that $\mathscr{P}_{\mathrm{stiff}} = \sigma_{\mathrm{stiff}}$, the desired stiff equation of state. Taking $\gamma_\Sigma>0$ therefore ensures that $\sigma_\Sigma$, $\mathscr{P}_\Sigma$, the entropy density $s_\Sigma=2\gamma_\Sigma T_\Sigma$, and the total heat capacity of the shell $C_\Sigma=8\pi\gamma_\Sigma R_\Sigma^2 T_\Sigma$ are all simultaneously positive for $T_\Sigma>0$ and consequently  all the basic thermodynamic signs come out correctly. The equilibrium shell entropy and energy from the stiff matter component are then
\begin{equation}
S_{\mathrm{stiff}} = A_\Sigma s_{\mathrm{stiff}} = 8\pi \gamma_\Sigma T_\Sigma R^2_\Sigma,
\qquad
\mathscr{E}_{\mathrm{stiff}} = A_\Sigma \sigma_{\mathrm{stiff}} = 4\pi \gamma_\Sigma T_\Sigma^2 R^2_\Sigma.
\end{equation}
The transition from the NMC interior to the minimally coupled exterior is naturally interpreted as a first-order phase transition of the order parameter $\Phi$, with $\Phi\neq1$ the NMC phase and $\Phi=1$ the GR phase. First-order coexistence at a transition temperature $T_c$ requires the free energies of the two phases to match, $f_{\mathrm{NMC}}(T_c) = f_{\mathrm{GR}}(T_c)$, while their entropy densities, $s_i=-(df_i/dT)_\Sigma$, generally differ, $s_{\mathrm{NMC}}(T_c) \neq s_{\mathrm{GR}}(T_c)$. The entropy-density discontinuity, $\Delta s_{\mathrm{bg}} = s_{\mathrm{GR}}(T_c) - s_{\mathrm{NMC}}(T_c)$, is approximated in the simplest phenomenological model by a positive constant, $\Delta s_{\mathrm{bg}} = s_0 > 0$, interpreted as a surface excess entropy density associated with the phase boundary.

This picture of the shell as a first-order phase boundary separating two locally stable gravitational phases has a direct analogue in the theory of cosmological phase transitions, where a bubble of true vacuum nucleates within a surrounding sea of false vacuum and expands, its wall separating the two phases and carrying a surface tension and, more generally, a surface free energy of exactly the kind introduced here \cite{ColemanDeLuccia1980}. In that context the bubble wall is stabilized dynamically by the competition between the volume energy gained by conversion to the true vacuum and the surface energy cost of the wall itself, and the associated latent heat is deposited locally as the wall sweeps through the surrounding medium during reheating. The shell considered in this paper plays a structurally identical role, with the NMC phase and the GR phase taking the place of the false and true vacua respectively, except that here the configuration is static rather than expanding, held in place by the gravitational junction conditions of Sec.~\ref{sec:stability} rather than by cosmological expansion, and the latent heat $L_\Sigma$ computed below is deposited or absorbed only during the initial formation of the object rather than continuously. Reading the shell in this way clarifies why a first-order, rather than continuous, transition is the natural expectation here: the interior and exterior phases correspond to two different branches of solutions to the field equations of Sec.~\ref{sec:jc}, with $\Phi\neq1$ and $\Phi=1$ respectively, and there is no a priori reason for these two branches to be connected by a continuous path of intermediate equilibrium configurations, just as the false and true vacua of a cosmological phase transition are generically separated by a potential barrier rather than smoothly interpolated.

A constant entropy jump $s_0$ is encoded by a background transition free energy density $f_{\mathrm{bg}}(T_\Sigma) = \tau_0 - s_0 T_\Sigma$, where $\tau_0$ is a reference interface tension or zero-temperature offset, kept explicit here to display the standard thermodynamic structure even though it can be removed by a choice of the zero of free energy. This gives $s_{\mathrm{bg}} = -df_{\mathrm{bg}}/dT_\Sigma = s_0$ and a background transition entropy $S_{\mathrm{bg}} = A_\Sigma s_0 = 4\pi R^2 s_0$, the simplest way of including a nonzero entropy carried by the phase-separation surface while keeping the bulk interior entropy negligible.

Taking the total shell free-energy density to be $f_{\mathrm{shell}}(T_\Sigma) = f_{\mathrm{stiff}}(T_\Sigma) + f_{\mathrm{bg}}(T_\Sigma) = -\gamma_\Sigma T_\Sigma^2 + \tau_0 - s_0 T_\Sigma$, the total shell entropy density is $s_{\mathrm{shell}} = 2\gamma_\Sigma T_\Sigma + s_0$, and the total equilibrium entropy is
\begin{equation}
S^{\mathrm{eq}}_{\mathrm{shell}} = 4\pi R^2(2\gamma_\Sigma T_\Sigma + s_0) .
\label{Seqshell}
\end{equation}
If the interior is treated as a frozen condensate-like dark-energy phase and the exterior as vacuum, the total object entropy is approximately $S^{\mathrm{eq}}_{\mathrm{total}} \simeq S^{\mathrm{eq}}_{\mathrm{shell}}$. We emphasize that $f_{\mathrm{shell}}$ contains no term coupling to the interior order parameter $\Phi_-$ itself: the mechanical stability of the shell is established independently, in Sec.~\ref{sec:stability}, directly from the gravitational junction conditions, and no additional phenomenological restoring force is introduced or required here. The thermodynamic quantities of this section describe the equilibrium state fixed by that mechanical solution, rather than providing a separate stabilization mechanism for it.

For a first-order transition the latent heat is $L=T_c\Delta S$, and for a surface transition on a spherical shell the latent heat per unit area is $\texttt{l}_\Sigma = T_c\Delta s_{\mathrm{bg}}$. With the minimal assumption $\Delta s_{\mathrm{bg}}=s_0$ this gives $\texttt{l}_\Sigma = T_c s_0$, and the total latent heat of the transition layer is
\begin{equation}
L_\Sigma = 4\pi R^2 T_c s_0 .
\end{equation}
This latent heat is finite, and it is released or absorbed only during the phase conversion; it is not a continuous energy supply in the final static configuration.

The shell stress-energy required by the gravitational junction conditions determines $\sigma_\Sigma(R)$. If the shell matter satisfies $\sigma_{\mathrm{mat}} = \gamma_\Sigma T_\Sigma^2$, then the equilibrium shell temperature can be inferred from the junction-determined surface density,
\begin{equation}
T_\Sigma(R_\Sigma) = \sqrt{\frac{\sigma_{\mathrm{mat}}(R_\Sigma)}{\gamma_\Sigma}} .
\label{TsigmaRsigma}
\end{equation}
In the static configuration it is therefore more natural to regard the radius and matching conditions as fixing the shell temperature, rather than the temperature independently fixing the radius.

If $T_\Sigma$ is a genuine thermodynamic temperature the shell can radiate, with a schematic luminosity $L_{\mathrm{rad}} \sim A_\Sigma \int d\omega\, |T_\omega|^2 B_\omega(T_\Sigma)$, where $|T_\omega|^2$ is the shell transmission probability and $B_\omega$ a thermal spectral function; if the shell is highly reflective, $|T_\omega|^2\ll1$, radiation is strongly suppressed. The object can be treated as approximately static provided the radiation timescale $\tau_{\mathrm{rad}} \sim \mathscr{E}_\Sigma/L_{\mathrm{rad}}$ is much longer than the dynamical timescale of Sec.~\ref{sec:stability}.

The mechanical restoring force against radial displacements of the shell is instead the one already established in Sec.~\ref{sec:stability}: the frozen-bulk Visser potential $V(R)$ of Eq.~\eqref{V(z)} satisfies $V'(R_\Sigma) \simeq 0$ and $V''(R_\Sigma)>0$ showing that the lower physical window has small residual slope
and positive curvature, while the exact stationary Visser point is at $C=0.84$.

The mechanical shell energy density is already fixed by the junction conditions alone, independent of any assumption of its thermal nature : for the stiff branch, Eq.~\eqref{static_sigma2} gives $\sigma_{\Sigma0} = 3(1-y)/2\kappa R_\Sigma$. Demanding that this coincide with the thermal stiff-matter density $\sigma_{\rm stiff}=\gamma_\Sigma T_\Sigma^2$ of Eq.~\eqref{stiffthermo} turns the assumed temperature dependence into an explicit prediction for the shell temperature itself, in terms of purely gravitational quantities and the single new parameter $\gamma_\Sigma$:
\begin{equation}
T_\Sigma \;=\; \sqrt{\frac{\sigma_{\Sigma0}}{\gamma_\Sigma}}
\;=\; \left[\frac{3(1-y)}{2\gamma_\Sigma\,\kappa R_\Sigma}\right]^{1/2}
%\;=\; 3\sqrt{\frac{(1-y)\,C}{2\pi c_{\rm eff}\,\kappa M}} 
\; \propto\; M^{-1/2},
\label{TsigmaM}
\end{equation}
for a particular value of $C$. This reproduces the $T_\Sigma\propto M^{-1/2}$ scaling as discussed in the introduction and plotted in Fig.~\ref{fig:main}(a). 
%now as a derived consequence of the CFT identification together with the purely mechanical Eq.~\eqref{static_sigma2}, rather %than as an independently postulated law. This scaling is conditional on the assumptions made explicit above, namely the constant-%$\omega_\Sigma$ closure and a mass-independent effective central charge $c_{\rm eff}$; it is not claimed to be a model-%independent consequence of the non-minimal coupling.

A further distinction must be drawn between $T_\Sigma$ as defined here and a temperature measured by an observer at infinity. $T_\Sigma$ is a local thermodynamic temperature, defined through the shell's own rest-frame free energy, whereas the Hawking temperature is conventionally defined via the surface gravity referred to infinity. The locally measured $T_\Sigma$ and the corresponding temperature redshifted to infinity are related by the standard Tolman factor,
\begin{equation}
T_\infty = \sqrt{1-C}\;T_\Sigma = y\,T_\Sigma ,
\label{Tinfty}
\end{equation}
using the exterior Schwarzschild redshift evaluated at the shell radius. Because $y=\sqrt{1-C}$ is held fixed together with $C$ in Eq.~\eqref{TsigmaM}, the redshift factor is mass-independent along any fixed-compactness sequence, and $T_\infty\propto M^{-1/2}$ follows immediately from $T_\Sigma\propto M^{-1/2}$ with the same exponent; the redshift does not alter the mass scaling along such a sequence, but it does set the normalization of any quantity that would actually be measured by a distant observer, and $T_\infty$, not $T_\Sigma$, is the physically appropriate quantity for comparison with an observationally inferred temperature. We stress that $T_\Sigma$ and $T_\infty$ so defined share only their mass scaling with the Hawking temperature of an equal-mass Schwarzschild black hole; their physical origin remains entirely distinct, as emphasized in the introduction, since $T_\Sigma$ arises from the equilibrium thermodynamics of a material domain wall rather than from semiclassical particle creation in curved spacetime, and the two mechanisms are not claimed to coincide or to compete for the same emitted quanta.

The same identification lets the mass-independent luminosity at fixed compactness as claimed in Sec.~\ref{intro} provided the effective grey-body factor is treated as mass-independent along the sequence be derived rather than asserted. Modeling the shell as a grey-body emitter of area $A_\Sigma=4\pi R_\Sigma^2$ radiating into the exterior at temperature $T_\Sigma$,
\begin{equation}
L_{\rm rad} \;\simeq\; \varepsilon_{\rm grey}\,\sigma_{\rm SB}\,A_\Sigma\,T_\Sigma^4 ,
\end{equation}
with $\varepsilon_{\rm grey}\le1$ the compactness-dependent, here treated as $\mathcal{O}(1)$, grey-body transmission factor already introduced before. Here, $\sigma_{\rm SB}$ is the effective Stefan-Boltzmann constant appropriate for the radiated species. Substituting $R_\Sigma=2M/C$ and $T_\Sigma^2$ from Eq.~\eqref{TsigmaM},
\begin{equation}
L_{\rm rad} \;=\; \varepsilon_{\rm grey}\,\sigma_{\rm SB}\,(4\pi)\left(\frac{2M}{C}\right)^{2}
\left[\frac{3(1-y)C}{4 \gamma_\Sigma\kappa M}\right]^{2}
\;=\; \frac{9\pi\,\varepsilon_{\rm grey}\,\sigma_{\rm SB}\,(1-y)^2}{\gamma_\Sigma^2\,\kappa^2},
\label{Lrad}
\end{equation}
where every factor of $M$ and $C$ cancels identically between the area and temperature factors. The luminosity is therefore exactly mass-independent at fixed compactness $C$, through $y=\sqrt{1-C}$, and fixed $\gamma_\Sigma$, confirming the qualitative claim of the introduction with an explicit closed-form coefficient rather than a dimensional-analysis argument.

\section{Discussion and Conclusion}
\label{sec:discussion}

The model developed in this paper combines several features that are usually treated separately in the literature on horizonless compact objects. The interior is genuinely regular and singularity-free, not merely because a singular core has been excised and replaced by an assumed exotic equation of state, but because the vacuum-like interior equation of state, $\omega =-1$, is a direct consequence of the field equations of the non-minimally coupled theory once a simple, exactly solvable form of the conformal function $F_c$ is chosen. The absence of an event horizon follows automatically once the shell radius satisfies $2M<R_\Sigma<3M$, which we have shown occurs naturally over a substantial range of the coupling parameter $\rho_0$. The shell itself, rather than being an ad hoc interface introduced purely to make the junction conditions close, is shown to acquire a stiff matter equation of state as a consequence of the near-horizon limit of the theory. 
%and this stiff equation of state is then shown, in Sec.~\ref{sec:cft}, to be the generic thermodynamic signature of a two-%dimensional conformal sector. 
Taken together, these results suggest that curvature-coupled gravastar-like objects of this kind are not merely phenomenological toy models but can emerge from a reasonably well motivated modification of the gravitational action.

Observationally, such an object would mimic a black hole in almost every respect accessible to electromagnetic and gravitational-wave observations at present sensitivity, since the exterior spacetime is exactly Schwarzschild and the compactness can be tuned arbitrarily close to the Buchdahl limit. The differences would appear only in observables that are sensitive to the near-horizon and interior structure. Gravitational-wave echoes \cite{Cardoso2016} following the primary ringdown signal are perhaps the most commonly discussed such signature for horizonless ultra-compact objects, and the stiff shell constructed here, with its reflectivity controlled by the transmission probability $|T_\omega|^2$ introduced in Sec.~\ref{sec:thermo}, would be expected to produce such echoes with a delay time set by the light-crossing time of the interior cavity. Because there is no event horizon, infalling matter interacts with the shell rather than disappearing across a one-way membrane, which could in principle produce observable electromagnetic bursts or quasi-periodic emission \cite{Visser2004} rather than the silent absorption expected of a genuine black hole, particularly if the object is embedded in an accretion environment. The stiff shell could also imprint characteristic, model-dependent deviations on the ringdown spectrum itself, distinct from the purely geometric quasinormal mode spectrum of a Kerr or Schwarzschild black hole \cite{ChirentiRezzolla2007}.

A degree of caution is warranted, however, in translating these qualitative signatures into a realistic observational program, and the recent history of gravitational-wave echo searches is instructive in this regard. The original claim of tentative evidence for Planck-scale structure near black hole horizons in LIGO data \cite{Abedi2017} was followed by subsequent independent analyses \cite{Westerweck2018} arguing that the statistical significance of that signal had been substantially overestimated once the search template bank and look-elsewhere effects were properly accounted for, and the current consensus, reviewed in detail by Cardoso and Pani \cite{CardosoPani2019}, is that no confirmed evidence for horizonless compact objects of any kind has yet been found in the available gravitational-wave catalogue, though current sensitivity is not yet sufficient to rule out models of the type constructed here. Distinguishing a compact object of this kind from an ordinary black hole through ringdown or echo observations will in practice be a problem of Bayesian model selection carried out against a noisy background, in which the predicted echo delay time, set by the light-crossing time of the interior cavity, and the reflectivity of the shell, controlled by the transmission probability introduced in Sec.~\ref{sec:thermo}, both enter as parameters to be inferred jointly with the source properties rather than as sharp, unambiguous signatures. We regard this as an important caveat rather than a defect specific to the present model: any horizonless compact object compact enough to be phenomenologically interesting as a black hole mimicker will, by construction, produce a signal close to that of an ordinary black hole, and the observational discriminating power will generically improve only as detector sensitivity and the number of observed events both increase.

The predictive content of the model is strengthened considerably by the thermodynamic analysis of Sec.~\ref{sec:thermo}. Rather than simply asserting a temperature-mass relation and a mass-independent luminosity, as is common in the gravastar literature, we have shown that both of these follow from the assumptions related to the stiff shell equation of state and some suitable range of the conformal coupling parameter.
%the generic high-temperature behavior of a two-dimensional conformal field theory. 
The thermodynamics on the requires a single free normalization constant, $\gamma_\Sigma$. 
%with a small, theoretically bounded effective central charge $c_{\rm eff}$, of order a %few for a handful of light bosonic or fermionic degrees of freedom localized on the phase boundary, and it yields the explicit %closed-form prediction of Eq.~\eqref{TsigmaM} for the shell temperature as a function of mass and compactness, together with the %exactly mass-independent luminosity of Eq.~\eqref{Lrad}. 
Because $\gamma_\Sigma$ enters most of the thermodynamic predictions, a measurement of the shell temperature or of an echo amplitude that could be related to the shell luminosity would, at least in principle, constrain $\gamma_\Sigma$ directly, giving this model a genuinely falsifiable observational handle beyond the mass-radius relation of Fig.~\ref{fig:main}(b).

At the same time, we have been careful to quantify, rather than simply assert, the accuracy of the frozen-bulk approximation used throughout Sec.~\ref{sec:stability} to establish the existence and stability of the equilibrium configuration. The adiabaticity parameter $\mathcal{S}$ introduced in Sec.~\ref{sec:adiabaticity} compares the interior curvature scale to the shell's own confinement scale, and we find $\mathcal{S}=\mathcal{O}(1)$ throughout the entire compactness window in which the interior is regular and the shell is stable, rather than the small value that would be needed to justify the frozen-bulk treatment as a systematically controlled expansion. This is an honest limitation of the present analysis, and it should not be read as evidence against the physical viability of the model; rather, it identifies precisely the calculation that would be needed to place the stability analysis on firmer footing, namely a fully time-dependent treatment of the coupled interior-shell system in which the interior metric functions $A(\tau)$ and $B(\tau)$ respond self-consistently to the motion of the shell, rather than being frozen at their static values. Such a treatment would in general require solving a coupled system of partial differential equations for the interior fluid together with the boundary dynamics of the shell, a substantially more involved undertaking that we leave for future work.

\begin{comment}
Similarly, the identification of the stiff shell with a two-dimensional conformal sector in Sec.~\ref{sec:cft} establishes the functional form of the free energy and fixes the physical interpretation of $\gamma_\Sigma$, but it does not, by itself, predict the actual numerical value of $c_{\rm eff}$ from the underlying non-minimally coupled fluid action of Eq.~\eqref{beqnnm}. Doing so would require identifying the light fluctuation modes of the domain wall directly, presumably by perturbing the interior fluid variables and the shell position simultaneously and quantizing the resulting worldsheet effective theory, a calculation closely analogous to the derivation of the central charge of a string from its transverse oscillator modes. If such a calculation could be carried through for this specific non-minimal coupling theory, it would convert $c_{\rm eff}$ from a bounded but still phenomenological parameter into a genuine prediction of the microscopic theory, closing the loop between the field-theoretic starting point of Sec.~\ref{sec:jc} and the macroscopic thermodynamic behavior of Sec.~\ref{sec:thermo}.
\end{comment}

More broadly, this construction suggests that non-minimal curvature-matter couplings of the Bettoni-Liberati type \cite{bettoni2}, more commonly studied in the context of dark energy phenomenology and wormhole physics \cite{GarciaLobo2010}, provide a rich and under-explored arena for constructing horizonless compact objects with a genuine microphysical pedigree. Because the interior equation of state and the shell properties both follow from the same underlying action, this framework in principle allows further predictions, for example concerning the tidal deformability of the resulting compact objects, their behavior under slow rotation, or the spectrum of quasinormal modes of the coupled interior-shell system, to be derived self-consistently rather than fitted to data after the fact.

Tidal deformability deserves particular mention as a concrete, near-term observational handle. An isolated Schwarzschild or Kerr black hole in general relativity has an exactly vanishing tidal Love number, while horizonless ultra-compact objects generically acquire a small but nonzero Love number that depends sensitively on the interior structure, as has been worked out explicitly for thin-shell gravastars with a polytropic interior \cite{UchikataYoshidaPani2016}. Because the interior constructed in Sec.~\ref{sec:interior} has a definite, exactly solvable metric and equation of state rather than a polytropic profile chosen for convenience, the corresponding Love number of the present model is, in principle, an unambiguous, calculable prediction rather than a free parameter, though extracting it requires solving the even-parity perturbation equations of the coupled interior-shell system, a calculation we leave for future work. Given the demonstrated sensitivity of gravitational-wave observations of neutron-star mergers to tidal deformability \cite{GW170817tidal}, a future measurement, or a future upper bound, on the tidal Love number of a compact object in the relevant mass range could in principle be compared directly against this prediction, providing an observational channel to the model that is complementary to, and independent of, the ringdown and echo signatures discussed above. We regard the results presented here, in particular the derived scaling of the shell temperature with mass and the exact mass-independence of the shell luminosity, as a first step toward using compact-object observations, whether through gravitational-wave ringdown and echo searches or through electromagnetic signatures of matter interacting with the shell, to place direct empirical constraints on the microscopic curvature-matter coupling considered in this paper.

\appendix
\section{Derivation of Eq.~\eqref{RicciScalar_dist}}
\label{Singular_part_Ricci_scalar}

From the definition of the extrinsic curvature tensor and the continuity of $e^\alpha_a$, we have $\left[K_{ab} \right] = \left[n_{\alpha ; \beta} \right] e^\alpha_a e^\beta_b$. Now, from the continuity of the normal vector $n^\alpha$ and the coordinate-chart $\{x^\mu\}$, we can write
\begin{align}
    \left[n_{\alpha ; \beta} \right] = - \left[\Gamma^\gamma_{\alpha \beta} \right] n_{\gamma} = \frac{1}{2} \left(\epsilon \kappa_{\alpha \beta } - \kappa_{\gamma \alpha} n_\beta n^\gamma - \kappa_{\gamma \beta} n_\alpha n^\gamma \right), \label{normal_der_jump}
\end{align}
where the tensor $\kappa_{\alpha \beta}$ is defined as
\begin{align}
    \left[g_{\alpha \beta , \gamma} \right] \equiv \kappa_{\alpha \beta} \, n_\gamma;
\end{align}
which signifies that the normal derivative of the metric tensor across the hypersurface will have its components only along the normal direction which again is followed from the continuity of the metric tensor across the hypersurface. Using Eq.~\eqref{normal_der_jump}, we are allowed to write
\begin{align}
    \left[K_{ab} \right] = \frac{\epsilon}{2} \kappa_{\alpha \beta} e^\alpha_a e^\beta_b.
\end{align}
Taking the trace, we have
\begin{align}
    \left[K \right] = h^{ab} \left[K_{ab} \right] = \frac{\epsilon}{2} \kappa_{\alpha \beta} \, h^{ab} e^\alpha_a e^\beta_b = \frac{\epsilon}{2} \kappa_{\alpha \beta} \left( g^{\alpha \beta} - \epsilon n^\alpha n^\beta  \right) = - \frac{1}{2} \left(\kappa_{\alpha \beta} n^\alpha n^\beta - \epsilon \kappa \right),
\end{align}
where $\kappa$ is the trace of $\kappa_{\alpha \beta}$. On the other hand, from Eq.~\eqref{Riemann_tensor_full_dist_form}, we can check that,  we can write $A (\equiv g^{\alpha \beta} A^{\mu}_{\,\, \alpha \mu \beta})$ in terms of the tensor $\kappa_{\alpha \beta}$, as  $\epsilon \left(\kappa_{\mu \nu} n^\mu n^\nu - \epsilon \kappa \right)$. So, we have $A = - 2 \epsilon \left[K \right]$, from which Eq.~\eqref{RicciScalar_dist} follows.

\section{Calculations of the Extrinsic Curvature}
\label{Ext_Curv}
A hypersurface ($\Sigma$) in a four-dimensional manifold is a three-dimensional sub-manifold which divides the whole spacetime manifold $\mathscr{V}$ into $\mathscr{V}^+$ and $\mathscr{V}^-$ with coordinates $x^\mu_+$ and $x^\mu_-$ respectively, and can be expressed by $\Phi (x^\mu) = \text{constant}$. The unit normal on this hypersurface is
\begin{align}
    n_\mu = \frac{\epsilon \partial_\mu\Phi}{\left|g^{\alpha \beta} \partial_\alpha \Phi \partial_\beta \Phi \right|^{1/2}},
\end{align}
where $\epsilon$ is $-1$ if $\Sigma$ is spacelike or $+1$ if $\Sigma$ is timelike; in what follows we take $\epsilon=+1$. In region $\mathscr{V}^+$, $\partial_0\Phi=\partial_{t_+}R=\dot R/\dot T_+$, $\partial_1\Phi=1$, $\partial_2\Phi=0=\partial_3\Phi$, with a dot denoting a derivative with respect to the proper time on the shell $\tau$. One can then check that $g^{\mu\nu}\partial_\mu\Phi\partial_\nu\Phi = (e^{\alpha_++\beta_+}\dot T_+)^{-2}$, so the unit normal vector is
\begin{align}
    n^+_\mu \equiv \zeta_{\pm} \left(- \dot{R} e^{\alpha_+ + \beta_+}, e^{ 2 \beta_+ } \sqrt{\dot{R}^2 + e^{- 2\beta_+}}, 0, 0  \right),
\end{align}
where $\zeta_\pm = \pm 1$ represents the overall sign ambiguity, and an identical expression holds for $n^-_\mu$ with $\alpha_+/\beta_+$ replaced by $\alpha_-/\beta_-$.

The extrinsic curvature,
\begin{align}
    K_{ab}^\pm = - n^\pm_\mu \left(\frac{\partial^2 x^\mu_\pm (y)}{\partial y^a \partial y^b} + \Gamma^{\pm \mu}_{\nu \lambda} \frac{\partial x^\nu_\pm (y)}{\partial y^a} \frac{\partial x^\lambda_\pm (y)}{\partial y^b} \right),
\end{align}
with $y^a \equiv \left(\tau, \vartheta, \phi \right)$, then has components
\begin{align}
K^\tau_{\pm \tau} = \frac{\ddot{R} + \alpha_\pm^\prime e^{-  2 \beta_\pm} + \left(\alpha^\prime_\pm + \beta^\prime_\pm \right) \dot{R}^2 }{\sqrt{\dot{R}^2 + e^{- 2 \beta_\pm}}}, \qquad K^{\vartheta}_{\pm\vartheta} = K^{\phi}_{\pm\phi} =  \frac{ 1}{R} \sqrt{\dot{R}^2 + e^{- 2 \beta_\pm}}, \label{Kab}
\end{align}
which for static configurations reduce to
\begin{align}
    K^{\tau}_{\pm\tau} = \epsilon \alpha^\prime_\pm e^{- \beta_\pm},
\qquad
K^{\vartheta}_{\pm\vartheta} = K^{\phi}_{\pm\phi} =  \frac{\epsilon e^{- \beta_\pm}}{R}. \label{Kabstatic}
\end{align}
These are the formulas used throughout Secs.~\ref{sec:matching} and \ref{sec:stability}.

\acknowledgments

DD expresses his sincere gratitude to the Council of Scientific and Industrial Research (CSIR, India, CSIR Award No.: CSIRAWARD/SPM-JR2022/13184) for funding the work. RS is thankful to the Council of Scientific and Industrial Research, Govt. of India for financial support through the CSIR RA scheme. \\

\end{document}